\documentclass[reprint,aps,prd,onecolumn,notitlepage,nofootinbib,
showpacs,preprintnumbers,superscriptaddress]{revtex4-1}
\usepackage[T1]{fontenc}
\usepackage[utf8]{inputenc}
\usepackage{bm}
\usepackage{amsmath}
\usepackage{amssymb}
\usepackage{esint}
\usepackage{color}
\usepackage{graphicx}
\PassOptionsToPackage{normalem}{ulem}
\usepackage{ulem}

\usepackage{times}
\usepackage{hyperref}
\usepackage{amsthm,bm,mathrsfs,stmaryrd,amsmath}

\newcommand{\nn}{\nonumber\\}
\usepackage[T1]{fontenc}
\usepackage{caption}
\usepackage{subcaption}
\usepackage{dsfont}
\usepackage{extarrows}
\usepackage[aligntableaux=center,boxsize=0.8em]{ytableau}

\allowdisplaybreaks[1]

\def\({\left(}
\def\){\right)}
\def\[{\left[}
\def\]{\right]}
\def\<{\left\langle}
\def\>{\right\rangle}

\def\be{\begin{eqnarray}}
\def\ee{\end{eqnarray}}
\def\nn{\nonumber\\}

\def\G{\Gamma}

\def\D{\Delta}

\def\nn{\nonumber\\}
\def\pa{\partial}

\begin{document}

\title{$\phi^n$ trajectory bootstrap
}

\author{Wenliang Li}
\affiliation{School of Physics, Sun Yat-Sen University, Guangzhou 510275, China}
\email{liwliang3@mail.sysu.edu.cn}

\begin{abstract}
We perform an extensive bootstrap study of Hermitian and non-Hermitian theories 
based on the novel analytic continuation of $\<\phi^n\>$ or $\<(i\phi)^n\>$ in $n$. 
We first use the quantum harmonic oscillator to illustrate various aspects of the $\phi^n$ trajectory bootstrap method, 
such as the large $n$ expansion, matching conditions, exact quantization condition, and high energy asymptotic behavior.  
Then we derive highly accurate solutions for the anharmonic oscillators with 
the parity invariant potential $V(\phi)=\phi^2+\phi^{m}$ and the $\mathcal{PT}$ invariant potential $V(\phi)=-(i\phi)^{m}$ 
for a large range of integral $m$, 
showing the high efficiency and general applicability of this new bootstrap approach. 
For the Hermitian quartic and non-Hermitian cubic oscillators, 
we further verify that the non-integer $n$ results for $\<\phi^n\>$ or $\<(i\phi)^n\>$ 
are consistent with those from the wave function approach. 
In the $\mathcal{PT}$ invariant case, the existence of $\<(i\phi)^n\>$ with non-integer $n$ allows us 
to bootstrap the non-Hermitian theories with non-integer powers, 
such as fractional and irrational $m$. 
\end{abstract}

\maketitle
\tableofcontents

\section{Introduction}
In recent years, the non-perturbative bootstrap approach to strong coupling physics has received considerable attention 
due to the seminal work on conformal field theory (CFT) \cite{Rattazzi:2008pe,Poland:2018epd}.  
In the nonperturbative bootstrap methods, 
one attempts to deduce nonperturbative predictions from the self-consistency of physical observables, 
together with some basic assumptions. 
The bootstrap approach to conformal theories is associated with the self-consistency constraints on the conformal correlators from the associativity of operator product expansion. 
For non-conformal theories, the correlators are less constrained due to the absence of conformal symmetry. 
Nevertheless, the non-conformal observables should also satisfy some self-consistent relations, 
which can be derived from a microscopic definition. 
\footnote{For conformal field theories, microscopic definitions are not necessary 
because they can be defined more axiomatically by the scaling dimensions and operator-product-expansion coefficients. } 
In the Lagrangian formulation, they are nothing but the Dyson-Schwinger equations for Green's functions \cite{Dyson:1949ha,Schwinger:1951ex,Schwinger:1951hq}, 
the quantum analogues of equations of motion. 
In the context of matrix model and gauge theory, they are also called loop equations \cite{Makeenko:1979pb, Makeenko:1980vm, Migdal:1983qrz} and have been revisited 
from the bootstrap perspective in 
\cite{Anderson:2016rcw,Lin:2020mme,Hessam:2021byc,Kazakov:2021lel, Kazakov:2022xuh,Cho:2022lcj, Li:2023nip}. 
In the Hamiltonian formulation, one can also derive some self-consistency constraints on the expectation values in (matrix) quantum mechanics \cite{Han:2020bkb}, 
which has been further explored in 
\cite{Han,Berenstein:2021dyf,Bhattacharya:2021btd,Aikawa:2021eai,Berenstein:2021loy,Tchoumakov:2021mnh,Aikawa:2021qbl,Du:2021hfw,Lawrence:2021msm,Bai:2022yfv,Nakayama:2022ahr,Li:2022prn,Khan:2022uyz,Morita:2022zuy,Berenstein:2022ygg,Blacker:2022szo,Nancarrow:2022wdr,Berenstein:2022unr,Lawrence:2022vsb,Lin:2023owt,Guo:2023gfi,Berenstein:2023ppj,Li:2023ewe,Fan:2023bld,John:2023him,Fan:2023tlh}. 

A general feature of the nonperturbative bootstrap approach is that 
the system is underdetermined, i.e., the number of equations is less than the number of unknowns. 
In quantum field theory, one would usually need an infinite number of constraints to solve the system, 
which is a formidable task. 
For example, a crossing constraint for a conformal correlator can involve an infinite number of scaling dimensions and operator-product-expansion coefficients. 
\footnote{In two-dimensional conformal field theory, the number of free parameters can be greatly reduced if they can be organized into a finite number of representations of the chiral algebra, such as the Virasoro algebra, 
which is associated with the null state condition. }
In this work, we will focus on the simpler case of $D=0+1$ dimension, i.e., quantum mechanics. 
For a monomial potential, the system of self-consistent equations may be solved up to a finite number of initial conditions. 
\footnote{However, there are infinitely many free parameters for an irrational power potential, 
which will be discussed in Sec.\ref{sec: irrational}. 
In multi-matrix models and lattice models, each Dyson-Schwinger equation may involve only a finite number of unknowns, 
but the number of undetermined parameters can grow with the number of loop equations under consideration, 
so the total number of free parameters can also be infinite. }
In most of the studies mentioned above, the indeterminacy problem was addressed 
by implementing numerical positivity constraints.   
A different resolution is to impose the null state condition 
 \cite{Li:2022prn,Li:2023nip,Guo:2023gfi,John:2023him}, 
which applies to non-positive systems 
and is closely related to the principle of minimal singularity  proposed more recently in \cite{Li:2023ewe}. 
Roughly speaking, one can determine the initial conditions by specific boundary conditions at infinity. 
\footnote{In \cite{Bender:2022eze,Bender:2023ttu}, 
the indeterminacy issue of the zero-dimensional Dyson-Schwinger equations \cite{Bender:1988bp} 
was resolved by the asymptotic behavior of the connected Green's functions. 
A main difference from \cite{Li:2023ewe} and the present work is 
that we derive the boundary conditions from the self-consistent equations. 
To the best of our understanding, the large $n$ asymptotic behaviors in \cite{Bender:2022eze,Bender:2023ttu} 
were deduced with some explicit input from the exact solutions. }

Let us give a brief summary of the bootstrap formulation of quantum mechanics. 
In $D=0+1$ dimension, we consider the Hamiltonian
\be
H=p^2+V(\phi)\,,
\label{Hamiltonian}
\ee
where $p, \phi$ are the momentum and position operators in quantum mechanics. 
We use the notation $\phi$, instead of $x$, because \eqref{Hamiltonian} can be viewed as the one-dimensional version of scalar field theory. 
A basic motivation for revisiting these one-body systems is that 
some nonperturbative bootstrap insights may also apply to the more challenging many-body systems. 
In this work, we consider the parity invariant potential 
\be
V(\phi)=\phi^2+\phi^{m}, 
\label{P-potential}
\ee
and the $\mathcal{PT}$ invariant potential
\footnote{Under the parity transformation, we assume that $\phi\rightarrow -\phi$, where $\phi$ is interpreted as the position operator or a pseudo-scalar. 
Under the time reversal transformation, we have $i\rightarrow -i$. 
Therefore, the building block $i\phi$ for the potential \eqref{PT-potential} is $\mathcal{PT}$ invariant. 
} 
\be
V(\phi)=-(i\phi)^{m}\,,
\label{PT-potential}
\ee 
where $m\geq 2$. 
For the parity invariant potential, we assume that $m$ is an even integer. 
For the $\mathcal{PT}$ invariant potential, we will consider both even and odd integers, 
as well as fractional and irrational $m$. 
These anharmonic oscillators can be viewed as the one-dimensional counterparts of 
the (multi-critical) Ising and Yang-Lee field theories. 
See \cite{Guo:2023qtt,Guo:2024bll} and references therein for recent perturbative bootstrap studies of 
the corresponding conformal field theories. 

The canonical commutation relation $[\phi,p]=i\hbar$ implies
that the commutator involving $\phi^n$ with integer $n$ is given by
\be
[p,\, \phi^n]=-i\hbar\,n\, \phi^{n-1}\,.
\label{commutator-p-xn}
\ee
In fact, we can analytically continue the integer parameter $n$ to complex numbers. 
Alternatively, it is easy to show that \eqref{commutator-p-xn} applies to non-integer $n$ 
using the position representation of the momentum operator $p=-i\hbar\, \pa_\phi$. 
For simplicity, we will focus on the expectation values associated with an eigenstate of the Hamiltonian \eqref{Hamiltonian} with real energy $E$. 
Assuming that the inner product is compatible with the symmetry of the Hamiltonian, we have
\be
\<H\mathcal O\>=E\<\mathcal O\>=\<\mathcal OH\>\,,
\ee
which implies
\be
\<[H,\phi^n]\>=\<[p^2,\phi^n]\>=0\,.
\ee
Together with \eqref{commutator-p-xn}, one can show that
\be
2i\hbar(n+2)_2\<\phi^{n+1}p\>+\hbar^2(n+1)_3\<\phi^n\>=0\,,
\label{expectation-commutator-H-xn}
\ee
where $(a)_b=\G(a+b)/\G(a)$ is the Pochhammer symbol. 
According to $\<[p^2+V(\phi),\phi^np]\>=0$, we further have
\be
-2i\hbar (n+2)_2 \<\phi^{n+1}p\>+4(n+3)\<\phi^{n+2}(E-V(\phi))\> -2\<\phi^{n+3}V'(\phi)\>=0\,,
\label{expectation-commutator-H-xnp}
\ee
where we have used $\<\mathcal O(p^2+V(\phi)-E)\>=0$ to eliminate the $p^2$ terms. 
Then the sum of \eqref{expectation-commutator-H-xn} and \eqref{expectation-commutator-H-xnp} gives
\be
\hbar^2(n+1)_3\<\phi^n\>+4E(n+3)\<\phi^{n+2}\>=4(n+3)\<\phi^{n+2}V(\phi)\>+2\<\phi^{n+3}V'(\phi)\>\,. 
\label{xn-recursion}
\ee 
Below we set $\hbar=1$. 
In the $\mathcal{PT}$ invariant theory with \eqref{PT-potential}, 
it is more natural to insert some $i$'s into the expectation values, i.e., $\<(i\phi)^n\>$. 
For a concrete potential in \eqref{P-potential} or \eqref{PT-potential}, 
one can see explicitly that \eqref{xn-recursion} leads to an underdetermined system of 
self-consistent equations for $\<\phi^n\>$. 
If one wants to determine a set of observables with small $n$ by \eqref{xn-recursion} with $m>2$, 
then their solutions are given by functions of the larger $n$ observables, 
which are new unknowns.
Furthermore, the energy $E$ is not fixed by these equations. 
Therefore, the system \eqref{xn-recursion} is underdetermined. 

In general, self-consistent equations imply that physical observables are closely related to each other. 
It is natural to package the set of self-consistent observables as a mathematical function, i.e., $G_n=\<\phi^n\>$. 
What kind of properties should be expected for the function $G_n$? 
Historically, the developments of the bootstrap approach to the strong interaction were deeply intertwined 
with analytic properties \cite{Regge:1959mz, Chew:book, Chew:1961ev}. 
In complex analysis, analyticity also implies some self-consistent constraints for complex functions.   
From a local perspective, the Cauchy-Riemann equations ensure that the complex derivative 
is given by a path-independent limit. 
\footnote{The existence of the first-order complex derivative in an open region 
implies that all the higher order derivatives exist, 
so complex differentiability is equivalent to analyticity.  } 
From a non-local perspective, 
the results of analytic continuations are path independent as long as there is no obstruction to path deformation. 
\footnote{Furthermore, the introduction of the complex infinity leads to a one-point compactification. 
}
Analyticity is both elegant and powerful. 
It should be useful to analytically continue the Green's functions $G_n$ to complex values of $n$ \cite{Li:2023ewe}. 
The complexification of $n$ is reminiscent of the complex angular momentum in Regge theory \cite{Regge:1959mz}. 
In analogy with the Regge trajectories, we can view $\phi^n$ as a novel type of analytic trajectories 
associated with the analytic continuation in $n$ that unifies different $G_n$.
\footnote{We may also define the analytic continuation in $n$ 
as a contour integral of the generating functional with a proper kernel. 
We thank Huajia Wang for suggesting this. 
}
In \cite{Li:2023ewe}, we proposed that the indeterminacy of bootstrap problems can be resolved by 
the principle of minimal singularity, 
i.e., the complexity of the singularity structure should be minimized. 
The basic idea is that the more physical solutions should be simpler than a generic self-consistent solution.  
To be more specific, we impose that the bootstrap solutions have a minimal number of asymptotic behaviors near the essential singularity at $n=\infty$. 
Note that the minimal boundary conditions may not be unique. 
The more physical choices depend on the system under consideration and the expected physical properties, such as vanishing constraints associated with 
the symmetry of the system.

We would like to emphasize that 
the principle of minimal singularity is different from the more well-known principle of maximal analyticity. 
The principle of maximal analyticity suggests that the domain of analyticity should be extended as far as possible, 
which mainly concerns the locations of singularities, 
but many functions can have the same analytic domain.  
The singularity structure is characterized by the locations and, crucially, the types of singularities. 
For example, it was already stated explicitly in \cite{Chew:1961ev} that ``maximal analyticity in linear momenta fails to
specify precisely the asymptotic behavior in momentum transfer''. 
\footnote{In \cite{Chew:1961ev}, Chew and Frautschi proposed to extend the principle of maximal analyticity to angular momentum, which is crucial to the bootstrap formulation. }
On the other hand, the principle of minimal singularity suggests that the singularity structures of physical solutions are less complicated, which concerns both the locations and types of the singularities. 
Let us take the extensions of the factorial function $n!$ as an example. 
The principle of maximal analyticity will prefer the Hadamard gamma function    
to the Bernoulli-Euler gamma function, 
as the former has no poles at non-positive $n$. 
In fact, the extensions with maximal analyticity are not unique. 
According to the large $n$ asymptotic behavior, the more standard Bernoulli-Euler gamma function is special in that it is minimally singular at the essential singularity $n=\infty$. 
\footnote{To be compatible with the factorial function, 
the large $n$ asymptotic behavior of a pseudogamma function can be different from that of the Bernoulli-Euler gamma function by oscillatory terms $\sin(\pi n)(\dots)$. }
In what sense is the essential singularity at $n=\infty$ more important than infinitely many poles at non-positive integers? 
As mentioned above, the essential singularity at $n=\infty$ can be viewed as a boundary of the more physical parameter space. 
Accordingly, the analytic function in $n$ is less oscillatory or, more generally, less complicated 
in the more physical region 
if the large $n$ asymptotic behavior is minimal. 

In this work, we will use the $\phi^n$ trajectory bootstrap to resolve the indeterminacy issue of \eqref{xn-recursion}.
Let us briefly summarize the basic idea of this new bootstrap approach. 
To solve for the Green's functions $\<\phi^n\>$,  
we perform the analytic continuation in $n$ and 
minimize the complexity of the singularity structure. 
As a function of $n$, we can study the minimally singular solutions for $\<\phi^n\>$ around two natural limits, i.e., $n=0$ and $n=\infty$. 
For relatively small $\text{Re}(n)$, we solve \eqref{xn-recursion} nonperturbatively 
and express the finite $n$ solutions in terms of a set of independent variables, i.e., initial conditions. 
\footnote{It is interesting to consider the small $n$ expansion, which seems less straightforward for the difference equations under consideration. }
For relatively large $\text{Re}(n)$, 
we solve \eqref{xn-recursion} perturbatively using the $1/n$ expansion and deduce accurate approximations for $\<\phi^n\>$ at finite $n$. 
\footnote{There may exist a semi-classical picture for the large $n$ limit 
as a saddle point approximation,  
which could be universal and related to black holes.  }
Then we impose the matching conditions
\footnote{We believe that this matching procedure can be extended to the more complicated many-body systems, 
which requires a better understanding of the large $n$ asymptotic behavior of these more challenging problems. 
A spectral representation may be useful for a large distance expansion based on the principle of cluster decomposition. } 
\be
\<\phi^n\>^\text{non-perturbative}=\<\phi^n\>^\text{perturbative}
\ee
in the overlap region around the matching order $n=M$,
which leads to accurate solutions to the underdetermined system  \eqref{xn-recursion}. 
In \cite{Li:2023ewe}, we have carried out this procedure successfully for the basic examples of the Hermitian quartic and non-Hermitian cubic oscillators. 
In this work, we would like to address two natural questions:
\begin{enumerate}
\item
Can we verify the results for $\<\phi^n\>$ or $\<(i\phi)^n\>$ at non-integer $n$ by a more standard method?

To the best of our knowledge, the Green's functions $\<\phi^n\>$ or $\<(i\phi)^n\>$ with non-integer $n$ have not been discussed before, 
so their precise values may seem irrelevant to the more physical Green's functions at integer $n$. 
In fact, there exist certain ambiguities in the minimally singular solutions, 
which are absent at integer $n$ due to $e^{2\pi i}=1$. 
It would be more reassuring if the minimally singular solutions at non-integer $n$ 
can be verified by a more standard approach. 

In this work, we will use the more standard wave-function formulation to verify the bootstrap solutions at non-integer $n$. 
According to the wave-function definitions in \eqref{Hermitian-inner-product} and \eqref{PT-inner-product}, 
the function $G_n$ is expected to be analytic when $\text{Re}(n)$ is in the range $(0, +\infty)$. 
The integration paths in \eqref{Hermitian-inner-product} and \eqref{PT-inner-product} are along the real axis.  
The integrals may diverge for $\text{Re}(n)<0$ due to the growth of the integrands 
around $\phi=0$. 
Then $G_n$ may have singularities at $\text{Re}(n)\leq 0$.
\footnote{A concrete analytic domain may extend to negative $\text{Re}(n)$. 
See Sec. \ref{sec:harmonic} for some explicit examples, which are expressed in terms of the gamma function. }
We will evaluate these integrals using \texttt{Mathematica}'s \texttt{NIntegral} with default settings except for the working precision,  
so the results are associated with the principal value of the exponential representation $e^{n\log \phi}$ for $\phi^n$.

\item
Can the $\phi^n$ trajectory bootstrap approach be efficiently applied to the anharmonic oscillators 
with higher powers or non-integer powers?

As the integral power $m$ increases, the self-consistent equations involve more initial conditions.  
A bootstrap procedure could cease to give accurate results if the computational complexity grows rapidly. 
The successes of the quartic and cubic examples do not guarantee that the higher power cases can be solved accurately with reasonable computational efforts. 
If a non-perturbative bootstrap method cannot solve the low dimensional problems accurately and efficiently, 
then it is unlikely that this can be of practical usefulness 
for the more complicated problems in higher dimensions. 

For the non-Hermitian $\mathcal {PT}$ invariant models 
\cite{Bender:1998ke, Bender:1999ek, Bender:2007nj,Bender:2010hf,r5,Bender:2023cem}, 
the case of non-integral power $m$ can also have a real and bounded-from-below energy spectrum. 
The self-consistent equations seem more subtle, 
as they explicitly involve $G_n=\<(i\phi)^n\>$ with non-integer $n$. 
It is interesting to consider these unconventional cases 
due to the connection to multi-critical Yang-Lee edge singularity \cite{vonGehlen:1994rp,Lencses:2022ira,Lencses:2023evr}. 
\footnote{The $D=2$ non-unitary minimal models $\mathcal M(2, 2n+3)$ with $n=1,2,3,\dots$ for the multi-critical Yang-Lee edge singularity proposed in \cite{Lencses:2022ira} are also related to $D=3$ non-unitary topological field theories \cite{Gang:2023rei}. 
As an extension of Zamolodchikov’s argument \cite{Zamolodchikov:1986db}, 
it was argued in \cite{Lencses:2022ira} that the Landau-Ginzburg description for these minimal models involves a non-canonical kinetic term or a fractional power interaction term. 
In a more recent work \cite{Lencses:2024wib}, it was proposed that their Landau-Ginzburg description is a field theory generalization of the $\mathcal {PT}$ invariant quantum mechanics \cite{Bender:1998ke} with integral power interaction terms. }
Furthermore, the analytic continuation in $m$  may lead to a new type of connected manifold 
associated with the multicritical Yang-Lee CFTs, i.e., a new kind of non-Hermitian conformal manifold that is based on the $\mathcal {PT}$ symmetry. 
In a broader context, the nonperturbative bootstrap solutions of the non-Hermitian systems may provide useful insights into 
the non-positive bootstrap studies in higher dimensions \cite{Gliozzi:2013ysa, Gliozzi:2014jsa, Li:2017agi, Hikami:2017hwv, Li:2017ukc,Li:2021uki,Li:2023tic}. 
\end{enumerate}

Let us give some general comments about the necessity of analytic continuation of $n$ to complex numbers. 
Strictly speaking, the extension from integral $n$ to complex $n$ is inevitable only for complex $m$. 
\footnote{It would be interesting to bootstrap the complex $m$ case, which is expected to be more subtle. }
For integral $m$, we can safely focus on the Green's functions $G_n$ with integral $n$,  
which avoids the potential ambiguities at non-integral $n$.   
Then the minimality of a bootstrap solution for $G_n$ is associated with the feasibility of 
an extension to a complex function in $n$ with minimal singularity at $n=\infty$. 
In both the integral $n$ and complex $n$ situations, we examine the number of different asymptotic behaviors at large $n$, 
so there seems no practical difference. 
However, the $n$ analytic continuation perspective leads to a unification of 
naively different branches of $G_n$, 
which is analogous to the Regge trajectory that led the bootstrap philosophy  \cite{Regge:1959mz, Chew:book, Chew:1961ev}. 
The explicit case of the cubic theory can be found in Fig. \ref{fig_ix-3}, where the naively 5 branches of $G_n$ are unified by the $n$ continuation. 
\footnote{In \cite{Li:2023ewe}, we have already discussed the merging phenomenon in the quartic theory. 
The unification in the cubic theory becomes more clear after we absorb the phases into the definition of $G_n$ by inserting some $i$'s, i.e., $G_n=\<(i\phi)^n\>$. 
A more general argument for considering complex $n$ is that 
quantum mechanics is fundamentally based on the complex number $i$.  
(See also the appendix of \cite{Bender:2023cem}.) }

As discussed above, we want to bootstrap some $\mathcal {PT}$ symmetric solutions in the non-integer $m$ cases, 
whose energy spectra are known to be real and bounded from below. 
For fractional $m$, the restriction of $G_n$ to integral $n$ is problematic for the bootstrap analysis 
because the recursion relation \eqref{xn-recursion} implies that they are related to those with non-integral $n$, 
then a minimal extension is given by a subset of rational $n$, depending on the specific fractional number $m$. 
For rational $m$, the discrete rotational symmetry of the recursion relation \eqref{xn-recursion} 
in \eqref{PT-rotation-symmetry-integral} 
leads to certain periodicity conditions on $G_n$, so the number of undetermined $G_n$ in a minimal extension is finite. 
\footnote{See the eqs. (7-9) in \cite{Li:2023ewe} for the general expressions for some exact solutions at $D=0$, 
where the explicit periodicity conditions are given in eq. (8). } 
The periodicity conditions in the fractional $m$ case can be viewed as a multi-fold covering version of the integral $m$ case. 
For instance, for $m=5/2$, the minimal extension is associated with half-integral $n$ and 
the periodicity condition for $G_1$ can be viewed as a two-fold covering version associated with $4\pi$ rotation, 
in analogy with the half-integer angular momentum arising from the double covering of SO(3). 
More details on the concrete example of $m=5/2$ can be found in Sec. \ref{sec:fractional-m}. 

In Sec.\ref{sec: irrational}, we will further consider some $\mathcal {PT}$ symmetric solutions 
in the irrational $m$ cases, 
which also have real and bounded-from-below energy spectra. 
For irrational $m$, the discrete rotational symmetry of the recursion relation \eqref{xn-recursion} does not 
give rise to periodicity conditions on $G_n$.  
Even in a minimal extension, there are infinitely many undetermined $G_n$ 
and the bootstrap solution is closer to the case of higher dimensional field theory.  
The irrational $m$ case is similar to incommensurability in the context of periodic or quasi-periodic systems,  
which arises from an irrational ratio of two periodicities. 

The rational $m$ case with a minimal set of $G_n$ can be viewed as certain decoupling limit of the generic situation. 
To make a connection with conformal field theory mentioned earlier, 
the irrational $m$ cases are similar in spirit to the irrational conformal field theories in two dimensions, 
which are different from the rational conformal field theories with finitely many independent parameters. 
An irrational $n$ extension of a rational $m$ model  
is also analogous to an irrational extension of a rational conformal field theory, 
which can encode additional nonlocal observables, 
such as connectivity properties.
 
The paper is organized as follows. 
In Sec. \ref{sec:harmonic}, we consider the basic example of the harmonic oscillator  
and present some explicit results of the bootstrap analysis both numerically and analytically. 
In Sec. \ref{sec:x2m}, we study the Hermitian parity invariant anharmonic oscillators. 
In the quartic oscillator example, 
we use the standard wave function formulation to compute $\<\phi^n\>$ at complex $n$ 
and show that the results are compatible with a parity-invariant minimally singular solution. 
We further solve the higher power oscillators accurately using the same matching procedure. 
In Sec. \ref{sec:ixm}, we investigate the non-Hermitian $\mathcal {PT}$-invariant anharmonic oscillators. 
We start with the integer $m$ cases, especially the cubic oscillator. 
Then we  extend the discussion to non-integer power $m$ by considering $\<(i\phi)^n\>$ with non-integer $n$, 
including the fractional and irrational $m$. 
In Sec. \ref{sec:discussion}, we summarize our results and discuss some directions for further investigations.

\section{The harmonic potential $V(\phi)=\phi^2$}
\label{sec:harmonic}
Let us consider the quantum harmonic oscillator as a basic example. 
We will discuss various aspects of the bootstrap analysis. 
In this simple example, 
some approximate numerical results can be promoted to exact analytic solutions. 

For the Hamiltonian $H=p^2+\phi^2$, the recursion relation \eqref{xn-recursion} reads
\be
(n+1)_3\,G_n+4E(n+3)\,G_{n+2}=4(n+4)\,G_{n+4}\,,
\label{Gn-recursion-harmonic}
\ee
where the normalization is set by $G_0=\<\phi^0\>=1$. 
The odd $n$ cases of $G_{n}$ vanish for parity symmetric solutions, 
so we can focus on the even $n$ cases. 
As a result, the recurrence relation \eqref{Gn-recursion-harmonic} is of ``second-order,'' 
which can be viewed as a discrete analog of 
the second-order differential equation for the wave function.

In the standard wave function approach, 
one first computes the general power series solution of the Schr\"{o}dinger equation.   
The series coefficients can be determined explicitly order by order. 
The second-order differential equation has two independent series coefficients. 
The large-$\phi$ asymptotic analysis shows that there are two possible types of leading asymptotic behaviors.
\footnote{In the differential equation, the $E$ term is subleading at large $\phi$, 
so the leading asymptotic behavior of the wave function is independent of $E$.
As in the standard procedure, we strip off an exponential part of 
the asymptotic behavior and focus on the remaining power series. }
To obtain a normalizable wave function, the divergent type should be absent, 
but it is associated with the typical large order behavior of the power series.
The matching between the finite order expressions and large order behavior 
implies that the power series should terminate. 
This is possible when $E$ takes some special discrete values 
\be
E=2k+1\,,
\label{HO-E}
\ee 
where $k$ is a non-negative integer. 
Then the power series solutions are given by the Hermite polynomials 
and the wave functions decay rapidly at large $\phi$. 
In this way, the energy of the harmonic oscillator is quantized by 
the normalizability assumption and the matching procedure. 

The steps of our bootstrap approach are in parallel to those in the wave function approach. 
At finite $n$, we can solve for $G_n$ one by one using the recursion relation \eqref{Gn-recursion-harmonic}. 
Some explicit examples are
\be
G_2=\frac 1 2E\,,\quad
G_4=\frac 3 8(E^2+1)\,,\quad
\ee
\be
G_6=\frac 5{16}E(E^2+5)\,,\quad
G_8=\frac{35}{128}(E^4+14E^2+9)\,,\quad
\ee
\be
G_{10}=\frac{63}{256}E(E^4+30E^2+89)\,,\quad
G_{12}=\frac{231}{1024}(E^6+55E^4+439E^2+225)\,.\quad
\ee 
For the harmonic potential, we find only one free parameter, i.e., $E$. 
Note that $G_{n}$ is an odd(even) function of $E$ 
when $n/2$ is an odd(even) integer.  

\begin{figure}[h]
	\centering
		\includegraphics[width=0.8\linewidth]{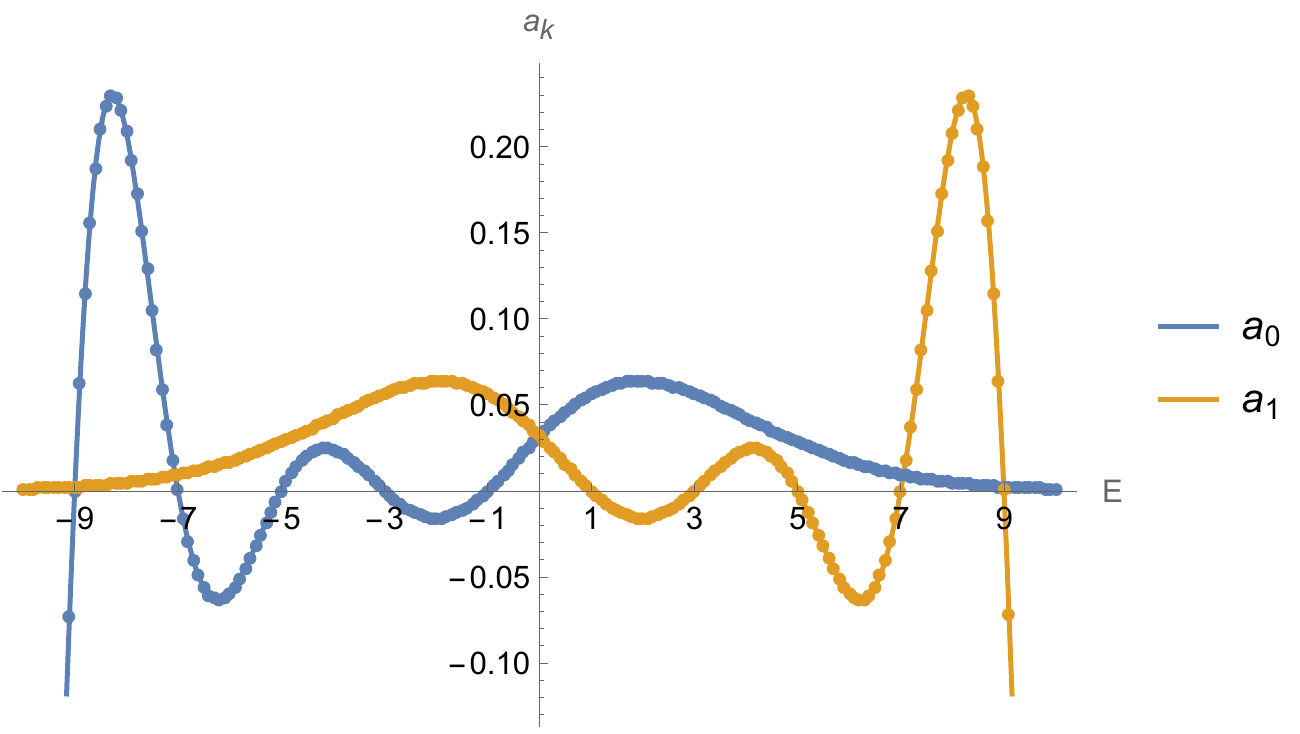}
	\caption{Prefactors of the two types of leading asymptotic behaviors in \eqref{HO-large-n-expansion} for the quantum harmonic oscillator at various $E$. 
	The bounded-from-below energy spectrum is associated with the zeros of $a_1[E]$. 
	The dots are obtained numerically by matching the non-perturbative solutions for $G_n$ with the perturbative $1/n$ series \eqref{HO-large-n-expansion}. 
	The curves are from the analytic expression for $a_0$ in \eqref{HO-a0-sol} and the relation between $a_0$ and $a_1$ in \eqref{a0a1-E}. 
	The numerical values are well interpolated by the analytic formulas. 
}
	\label{a0a1}
\end{figure}

To determine $E$, we study the asymptotic behavior of $G_n$ at large $n$,  
which can be derived from the dominant terms in \eqref{Gn-recursion-harmonic}
\be
n^3G_n\sim 4nG_{n+4}\quad (n\rightarrow \infty)\,.
\ee
The $E$ term is subleading due to the growth of $G_n$ in $n$. 
There are two possible types of leading asymptotic behaviors for integral $n$
\be
G_n\sim \frac{1+(-1)^n}{2}\,2^{n/2}\left[\G\left(\frac n 4\right)\right]^2\,\left(a_0+a_1\,(-1)^{n/2}\right)
\quad (n\rightarrow \infty)\,,
\ee
where we have imposed $G_{n}=0$ for odd $n$. 
If we further take into account the subleading terms in \eqref{Gn-recursion-harmonic}, 
we obtain the additional factors $n^{\frac{1+E}2}$, $n^{\frac{1-E}2}$. 
The subleading asymptotic behaviors are encoded in the $1/n$ series
\footnote{If we replace the leading behavior $2^{n/2}[\G(n/4)]^2n^{1/2}$ 
with $(1/2)_{\frac n 2}\,n^{-1/2}$, 
the large $n$ expansion coefficients take the form $c_{0,j}[E]\propto (\frac{1-E}{2})_j$. 
Then the $1/n$ series of the minimally singular solution terminates precisely at the exact values in \eqref{HO-E}, 
which is similar to the power series solutions associated with the normalizable wave functions. } 

\be
G_n\sim \frac{1+(-1)^n}{2}\,2^{n/2}\left[\G\left(\frac n 4\right)\right]^2
\sum_{j=0}^N\left(a_0\,{c_{0,j}}\,n^{\frac{1+E}2-j}
+a_1\,{c_{1,j}}\,(-1)^{\frac n 2}\,n^{\frac{1-E}2-j}\right)
\,,\quad
\label{HO-large-n-expansion}
\ee
where $N$ denotes the truncation order of the $1/n$ series. 
Note that  $a_k=a_k[E]$ and $c_{k,j}=c_{k,j}[E]$ are functions of $E$. 
We set $c_{0,0}=c_{1,0}=1$, 
so $(a_0, a_1)$ are fixed by the normalization condition $G_0=1$. 
The relative series coefficients can be solved systematically using \eqref{Gn-recursion-harmonic}. 
Some explicit results are
\be
c_{0,1}=\frac {1}{4} (2 E-5)\,,\quad
c_{0,2}=\frac{1}{96} (4 E^3- 12 E^2 - 40 E+75 )\,,
\ee
\be
c_{0,3}= \frac{1}{384} (8 E^4 - 92 E^3+ 292 E^2 - 238 E +147)\,,
\ee
\be
c_{0,4}=\frac{1}{92160} (80 E^6  - 
   1344 E^5+ 6320 E^4 - 1800 E^3 - 42040 E^2 + 86304 E -72225 )\,.\quad
 \ee
Since the recursion relation \eqref{Gn-recursion-harmonic} is invariant under the transformation
\be
G_n\rightarrow  (-1)^{n/2}\,G_n\,, \quad
E\rightarrow -E\,,
\ee
the two types of coefficients are related by
\be
c_{1,j}[E]=c_{0,j}[-E]\,,
\label{c0c1-E}
\ee
which can also be noticed from the explicit solutions. 
This is a discrete analog of the Symanzik/Sibuya rotation \cite{Sibuya}, 
which also appears in the anharmonic oscillators with higher powers. 
According to the normalization condition $G_0=1$, we further have
\be
a_1[E]=a_0[-E]\,.
\label{a0a1-E}
\ee
For a given $E$, we can extract the precise numerical values of $(a_0, a_1)$ 
by matching the finite $n$ solutions of $G_n$ with the $1/n$ series \eqref{HO-large-n-expansion}. 
In Fig. \ref{a0a1}, we present the results for $(a_0, a_1)$ in the range $-10\leq E\leq 10$. 
We can see that $a_1$ vanishes around $E=1,3,5,7,9$, 
which is in accordance with the exact solutions in \eqref{HO-E}. 

The fact that the exact solutions are related to the zeros of $a_1[E]$ can be explained 
by the principle of minimal singularity. 
The general solution in \eqref{HO-large-n-expansion} has two types of singular behaviors at $n=\infty$. 
To minimize the complexity of the singularity structure, 
we have two choices: $a_0=0$ or $a_1=0$.  
As the former has no solutions at large $E$, 
the quantization condition for a bounded-from-below energy spectrum is associated with the latter
\footnote{The other choice $a_0=0$ is associated with a bounded-from-above energy spectrum.}
\be
a_1[E]=0\,.
\label{HO-quantization-condition}
\ee
In this way, we determine the large $n$ asymptotic behavior up to a prefactor  
by the principle of minimal singularity and a spectral assumption, 
in analogy with the normalizability assumption in the wave function approach. 
\footnote{The choice of only one type of leading asymptotic behaviors for the wave function 
can also be viewed as a kind of minimal singularity assumption.} 

\begin{figure}[h]
	\centering
		\includegraphics[width=0.7\linewidth]{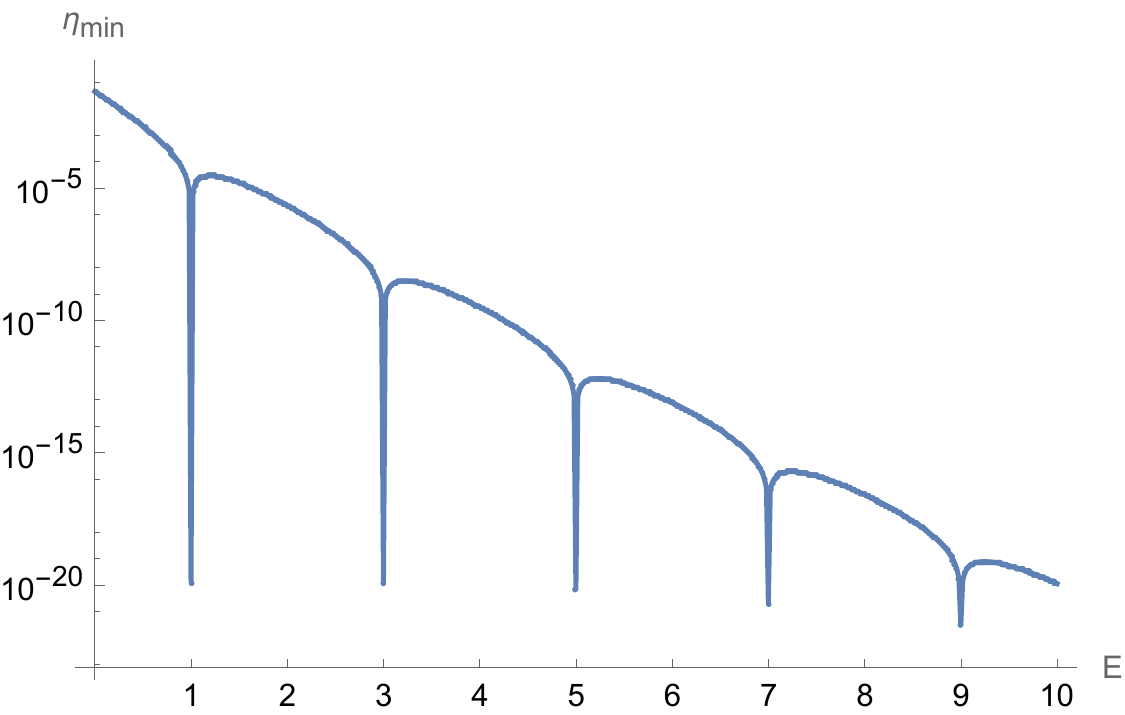}
	\caption{$\eta_\text{min}$ landscape of the quantum harmonic oscillator. 
	The local minima are consistent with the exact spectrum in \eqref{HO-E}.}
	\label{HO-eta-landscape}
\end{figure}

Let us use the quantization condition \eqref{HO-quantization-condition} to deduce the energy spectrum. 
We impose that the finite $n$ solutions for $G_n$ at relatively large $n$ match with the $1/n$ expansion of $G_n$. 
Since $a_1=0$, there remain two free parameters, i.e., $(E, a_0)$,  
which can be determined by two matching conditions
\be
G_M^{(\text{n.p.})}=G_M^{(\text{p.})}\,,\quad
G_{M+2}^{(\text{n.p.})}=G_{M+2}^{(\text{p.})}\,.
\label{matching-HO}
\ee
Note that $M$ denotes the matching order, 
$G_n^{(\text{n.p.})}$ indicates the non-perturbative finite $n$ expressions for $G_n$ from the recursion relation \eqref{Gn-recursion-harmonic}, 
and $G_n^{(\text{p.})}$ is given by the perturbative $1/n$ series in \eqref{HO-large-n-expansion} with $a_1=0$. 
As $(M, N)$ increase, 
it requires some efforts to deduce 
all the solutions of the matching conditions \eqref{matching-HO} 
due to the nonlinear $E$ dependence. 
Since we are mainly interested in the positive real energy solutions, 
we can reformulate the difficult problem of solving a set of highly nonlinear equations 
as the simpler least-squares problem. 
To measure the errors in the matching conditions \eqref{matching-HO}, 
we introduce the $\eta$ function 
\be
\eta=\sqrt{\sum_{n}\left(G_n^{(\text{n.p.})}-G_n^{(\text{p.})}\right)^2}\,,
\label{eta-HO}
\ee
where $n$ runs over the indices of the matching conditions. 
According to \eqref{matching-HO}, we choose $n=M, M+2$ for the harmonic oscillator.  
It is useful to divide $G_n$ by the leading asymptotic behavior for large $n$ so that each term is of order $\mathcal O(1)$. 
As $G_n^{(\text{p.})}$ is linear in $a_0$,  
it is straightforward to minimize the $\eta$ function for a given $E$. 
We can scan the $\eta_\text{min}$ landscape as a function of $E$. 
The solutions of the matching conditions \eqref{matching-HO} 
are associated with the local minima with $\eta_\text{min}=0$. 
In Fig. \ref{HO-eta-landscape}, we present the $\eta_\text{min}$ landscape for $(M,N)=(100,10)$, 
which contains 5 local minima with $\eta_\text{min}=0$ in the range $0<E<10$. 
All of them can be identified with the exact values from \eqref{HO-E},  
\be
E^{(k=0)}_{\text{approx.}}\approx 1+3\times 10^{-17}\,,\quad 
E^{(k=1)}_{\text{approx.}}\approx 3-3\times 10^{-13}\,,\quad 
E^{(k=2)}_{\text{approx.}}\approx 5+8\times 10^{-10}\,,\nonumber
\ee
\be 
E^{(k=3)}_{\text{approx.}}\approx 7-8\times 10^{-7}\,,\quad
E^{(k=4)}_{\text{approx.}}\approx 9+3\times 10^{-4}\,,
\ee
where the errors grow with $k$. 
There are more local minima at larger $E$.  
The solutions for $a_0$ are approximately given by
\be
4\sqrt 2\, \pi\, \G\left(\frac{1+E}{2}\right)a_0\approx 1+1\times 10^{-18}\,,
\ee
which indicates the analytic expression
\be
a_0[E]=\frac{1}{4\sqrt 2 \,\pi\, \G\left(\frac{1+E}{2}\right)}\,. 
\label{HO-a0-sol}
\ee
As $(M,N)$ increase, 
the numerical solutions converge rapidly to the analytic values in \eqref{HO-E} and \eqref{HO-a0-sol}. 
According to \eqref{a0a1-E} and \eqref{HO-a0-sol}, 
the explicit expression of the quantization condition \eqref{HO-quantization-condition} reads
\be
a_1[E]=\frac{1}{4\sqrt 2 \,\pi\, \G\left(\frac{1-E}{2}\right)}=0\,.
\label{HO-quantization-condition-explicit}
\ee
The corresponding solutions for $E$ are identical to the exact values in \eqref{HO-E}.

Let us make some consistency checks. 
When $E$ takes an exact value in \eqref{HO-E},  
we verify that the corresponding $1/n$ series is compatible with 
the minimally singular form, i.e., \eqref{HO-large-n-expansion} with $a_1=0$. 
For notational simplicity, we will not write the factor $\frac{1+(-1)^n}{2}$ explicitly. 
Some examples are
\be
G^{(E=1)}_n&=&\left(\frac1 2\right)_{n/2}
\sim 2^{n/2}\left[\G\left(\frac n 4\right)\right]^2n^{\frac{1+1}{2}}
\frac{1-\frac 3 {4 n}+\frac 9 {32n^2}+\frac{39}{128n^3}+\dots}{4\sqrt 2\,\pi\,\G(\frac{1+1}{2})}\,,
\ee
\be
G^{(E=3)}_n&=&\left(\frac 3 2\right)_{n/2}
\sim 2^{n/2}\left[\G\left(\frac n 4\right)\right]^2n^{\frac{1+3}{2}}
\frac{1+\frac 1 {4 n}-\frac {15} {32n^2}+\frac{75}{128n^3}+\dots}{4\sqrt 2\,\pi\,\G(\frac{1+3}{2})}\,,
\ee
\be
G^{(E=5)}_n&=&\frac{n^2+2n+2}{2}\left(\frac 1 2\right)_{n/2}
\sim  2^{n/2}\left[\G\left(\frac n 4\right)\right]^2{n^{\frac{1+5}{2}}}
\frac{1+\frac 5 {4 n}+\frac {25} {32n^2}-\frac{81}{128n^3}+\dots}{4\sqrt 2\,\pi\,\G(\frac{1+5}{2})}\,,\qquad
\ee
\be
G^{(E=7)}_n&=&\frac{n^2+2n+6}{6}\left(\frac 3 2\right)_{n/2}
\sim 2^{n/2}\left[\G\left(\frac n 4\right)\right]^2{n^{\frac{1+7}{2}}}
\frac{1+\frac 9 {4 n}+\frac {193} {32n^2}+\frac{147}{128n^3}+\dots}{4\sqrt 2\,\pi\,\G(\frac{1+7}{2})}\,,
\qquad  
\ee
where $\dots$ indicates higher order terms in the $1/n$ expansion. 
The concrete values of $a_0$ are also consistent with the analytic expression \eqref{HO-a0-sol}. 
If $E$ is not a positive odd integer, the large $n$ expansion should involve a nonzero $a_1$. 
In the simple case of $E=0$, 
the recursion relation \eqref{Gn-recursion-harmonic} can be solved explicitly
\be
G^{(E=0)}_n&=&\frac{(1+(-1)^{\frac n 2})\G(\frac{n+1}{2})\G(\frac{n+2}4)}{2\pi\G(\frac{n+4}4)}
\sim 
2^{n/2}\left[\G\left(\frac n 4\right)\right]^2\,{n^{\frac 1 2}}
\frac{1+(-1)^{\frac n 2}+\dots}{4\sqrt 2\, \pi\,\G(\frac{1+0}{2})}\,. \quad
\ee
Therefore, we have $a_1=a_0$ as expected from the $E\rightarrow -E$ symmetry. 
The exact value of $a_0$ at $E=0$ also confirms the analytic expression in \eqref{HO-a0-sol}.
For other even $E$, the solutions for $G_n$ are not invariant under the transformation $E\rightarrow -E$, 
but they can also be solved in closed form. 
For example, the solution for $E=2$ reads
\be
G_n^{(E=2)}&=&\frac{1-(-1)^{\frac n 2}}{2}\frac{4\G(\frac n 4+1)\G(\frac{n+1}{2})}{\pi\,\G(\frac n 4+\frac 1 2)}
+\frac{1+(-1)^{\frac n 2}}{2}\frac{2\G(\frac n 4+\frac 1 2)\G(\frac{n+3}{2})}{\pi\,\G(\frac n 4+1)}
\nn
&\sim& 
2^{n/2}\left[\G\left(\frac n 4\right)\right]^2\,
\left(
\frac{n^{\frac{1+2}{2}}(1-\frac 1 {4n}+\dots)}{4\sqrt 2\, \pi\, \G(\frac {1+2}{2})}
+(-1)^{\frac n 2}\,
\frac{n^{\frac{1-2}{2}}(1-\frac 9 {4n}+\dots)}{4\sqrt 2\, \pi\, \G(\frac {1-2}{2})}
\right)\,,\quad
\ee
which is also compatible with the general form of the $1/n$ series in \eqref{HO-large-n-expansion} 
and the analytic expressions for $(a_0, a_1)$. 

As we have deduced the complete energy spectrum, it is interesting to examine the large $E$  asymptotic behavior. 
For small $n$, the Green's functions can be approximated by
\be
G_n\sim E^{\frac n 2}\frac{(1/2)_{n/2}}{(1)_{n/2}}
\left(1+\frac {(n-2)_3}{24E^{2}}+\dots\right)\quad(E\rightarrow \infty)\,,
\ee
where $\dots$ denotes subleading terms at large $E$. 
We can also consider the large $n$ region. 
A resummation of the large $E$ contributions in the $1/n$ series leads to 
\footnote{The $1/n$ series of the anharmonic oscillators also allow for 
similar resummations of the large $E$ expansion. }
\be
\sum_{j}{c_{0,j}}\,n^{-j}
\sim e^{\frac{1}{24}\big(\frac{E^{3/2}}{n}\big)^2}
\Bigg(1+\frac{\frac{E^{3/2}}{n}}{2}\frac {1}{E^{1/2}}
-\Bigg(\frac {\big(\frac{E^{3/2}}{n}\big)^2} 8 +\frac {3\big(\frac{E^{3/2}}{n}\big)^4} {320}\Bigg)\frac{1}{E}+\dots
\Bigg)\,,\quad
\ee
which contains an exponential term. 
Alternatively, this can be computed systematically from the double expansion in $1/{E^{1/2}}$ and $ {E^{3/2}}/{n}$. 
The prefactor of the large $n$ series is given by
\be
a_0[E]\sim \frac {(2e)^{E/2}E^{-E/2}}{8\pi^{3/2}}
\left(1+\frac 1{12E}+\frac 1 {288 E^2}+\dots\right)\quad(E\rightarrow \infty)\,.
\ee
The vanishing condition \eqref{HO-quantization-condition-explicit} for the other prefactor  becomes
\be
a_1[E]\sim \frac {(2e)^{-E/2}E^{E/2}}{4\pi^{3/2}}\cos\Big(\frac \pi 2\,E\Big)
\left(1-\frac 1{12E}+\frac 1 {288 E^2}+\dots\right)=0\quad(E\rightarrow \infty)\,, 
\label{HO-quantization-condition-asymptotic} 
\ee
which gives the exact spectrum \eqref{HO-E} 
due to the exact form of the oscillatory part $\cos\big(\frac \pi 2\,E\big)$. 
In analogy with the WKB method, 
one should be able to derive the asymptotic quantization condition \eqref{HO-quantization-condition-asymptotic}  directly 
from the global asymptotic solution for $G_n$ at large $E$. 

It is also interesting to consider the composite operators involving the momentum operator $p$ or time derivatives. 
A simple example for multiple $p$ is
\be
\<p^n\>=\<\phi^n\>\,,
\ee
as the harmonic oscillator Hamiltonian and 
the canonical commutation relation are invariant under  
$p\rightarrow -\phi$, $\phi\rightarrow p$. 
A simple example for multiple time derivatives is
\be
\<\phi \frac{d^n\phi}{dt^{n}}\>=\pa_{t_2}^n\,G_2(t_1,t_2)\big|_{t_1\rightarrow t_2}=\frac{(2i)^{n}}{2}\left(\frac {1+(-1)^n}{2}E+\frac {1-(-1)^n}{2}\right)\,,
\ee
where $t_1,\,t_2$ are the time coordinates for the 2-point Green's function $G_2$ 
and we assume $t_1\geq t_2$. 
For simplicity, we assume that $n$ is a non-negative integer. 
A proper analytic continuation to complex $n$ requires some care. 
\footnote{A non-integer power of the differentiation operator involves the non-local properties of a function, 
which is different from the standard cases with integer powers. 
The generalization of derivatives and integrals to non-integer orders 
is known as fractional calculus. 
A basic example found by Lacroix is 
\be
\frac{d^n}{dx^n} x^a=\frac{\G(a+1)}{\G(a+1-n)}x^{a-n}\,,
\ee
where $a>0$. This formula can be derived from the Riemann–Liouville integral. 
Note that a fractional derivative of the constant function does not need to be zero. 
It may be interesting to consider higher derivative kinetic terms, 
which may have non-integer powers in the time derivative. }

Below we show that  the $\phi^n$ trajectory bootstrap method applies to the anharmonic oscillators 
that do not admit simple analytic solutions. 

\section{The parity invariant potential $V(\phi)=\phi^2+\phi^{m}$}
\label{sec:x2m}
In this section, we consider the Hermitian anharmonic oscillator
\be
H=p^2+\phi^2+\phi^m\,,
\label{H-xm}
\ee
where we assume that $m\geq 4$ and $m$ is an even integer.
The Hamiltonian \eqref{H-xm} is invariant under the parity transformation
\be
\phi\rightarrow -\phi\,.
\ee 
The recursion relation \eqref{xn-recursion} reads
\be
(n+1)_3G_n+4E(n+3)G_{n+2}=4(n+4)G_{n+4}+2(2n+m+6)G_{n+m+2}\,, 
\label{recursion-x2m}
\ee
where the Green's functions $G_n$ are defined as the expectation values
\be
G_n=\<\phi^n\>\,.
\ee
The normalization is fixed by $G_0=1$. 
For parity symmetric solutions, $G_n$ vanishes if $n$ is an odd integer. 
For $G_n$ with integer $n>0$, there are $m/2$ free parameters. 
We choose the independent set of free parameters as 
\be
(E\,,G_2\,, G_4\,,\dots, G_{m-2})\,.
\label{x-2m-free-parameters}
\ee
The other $G_{n}$ at integer $n$ can be determined by the recursion relation \eqref{recursion-x2m}.  
As $n$ increases, the analytic expressions of the nonperturbative solutions for $G_n$ are of high degree in $E$, but at most linear in $(G_2\,, G_4\,,\dots, G_{m-2})$. 

\subsection{$m=4$}
For the quartic oscillator $m=4$, the Hamiltonian reads
\be
H=p^2+\phi^2+\phi^4\,.
\label{quartic-H}
\ee
There are only two free parameters
\be
(E,\,G_2)\,.
\ee
The large $n$ expansion of the standard parity symmetric solution reads \cite{Li:2023ewe}
\be
G_n\sim 
a_0\frac{1+e^{2\pi i\frac n 2}}{2}3^{n/3}n^{1/6}\left[\G\left(\frac n 6\right)\right]^2 e^{-\left(\frac n 2\right)^{1/3}}
\left(1+\sum_{j=1}^{3N}c_j\left(\frac n 2\right)^{-j/3}\right)
\quad (n\rightarrow \infty)\,,\quad
\label{quartic-large-n-expansion}
\ee
where $N$ is the truncation order of the $1/n$ series. 
The explicit expressions of some low order coefficients are
\be
c_1= -E-\frac 1 6\,,\quad
c_2=\frac{36E^2+12E-11}{72}\,,\quad
c_3=\frac{-216 E^3-108E^2+198E-883}{1296}\,. 
\ee
The free parameters can be determined to high accuracy 
by the matching procedure. 
For example, the ground state solution corresponds to
\be
a_0&=& 0.484173090557323742122230381577...\,,
\label{quartic-a0}
\\
E&=& 1.39235164153029185565750787661...\,,
\label{quartic-E0}
\\
G_2&=& 0.305813650717587136934033799352...\,,
\label{quartic-G2}
\ee
which can be obtained from the matching conditions 
\be
G_M^{(\text{n.p.})}=G_M^{(\text{p.})}\,,\quad
G_{M+2}^{(\text{n.p.})}=G_{M+2}^{(\text{p.})}\,,\quad
G_{M+4}^{(\text{n.p.})}=G_{M+4}^{(\text{p.})}\,
\label{matching-x4}
\ee
with $(M,N)=(200,20)$. 
They are consistent with the diagonalization results. 

\begin{figure}[h]	
	\begin{subfigure}{0.47\textwidth}
	\raggedright
		\includegraphics[width=1\linewidth]{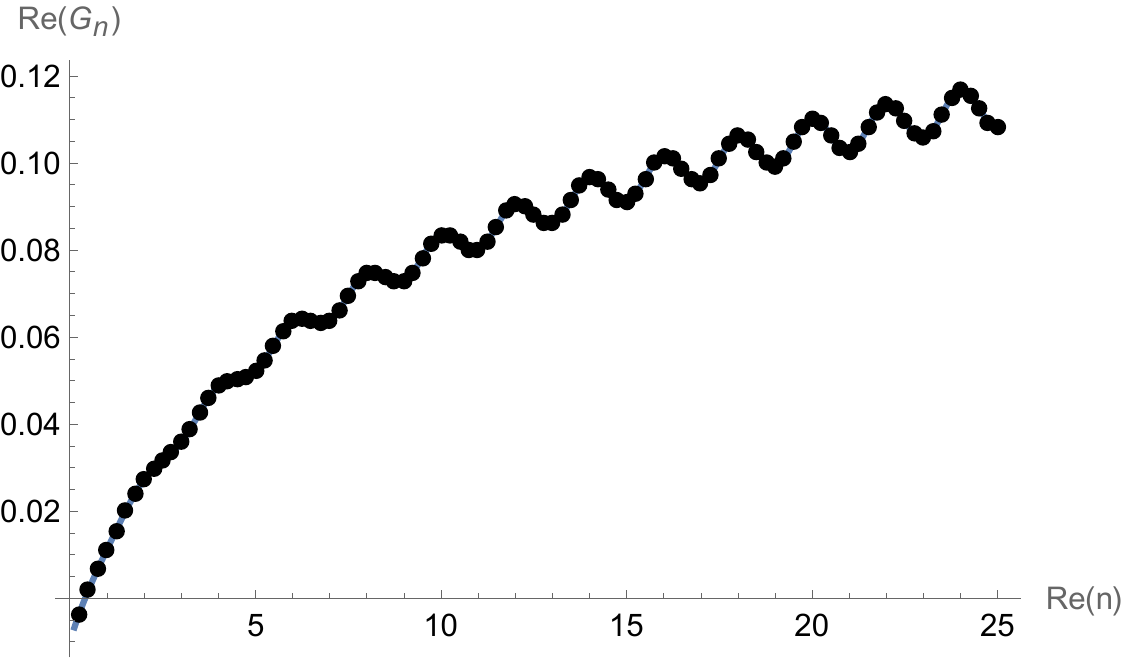}
		\caption{$\text{Re}(G_n)$ with $\text{Im}(n)=1$}
	\end{subfigure}
	\qquad
	\begin{subfigure}{0.47\textwidth}
	\raggedright
		\includegraphics[width=1\linewidth]{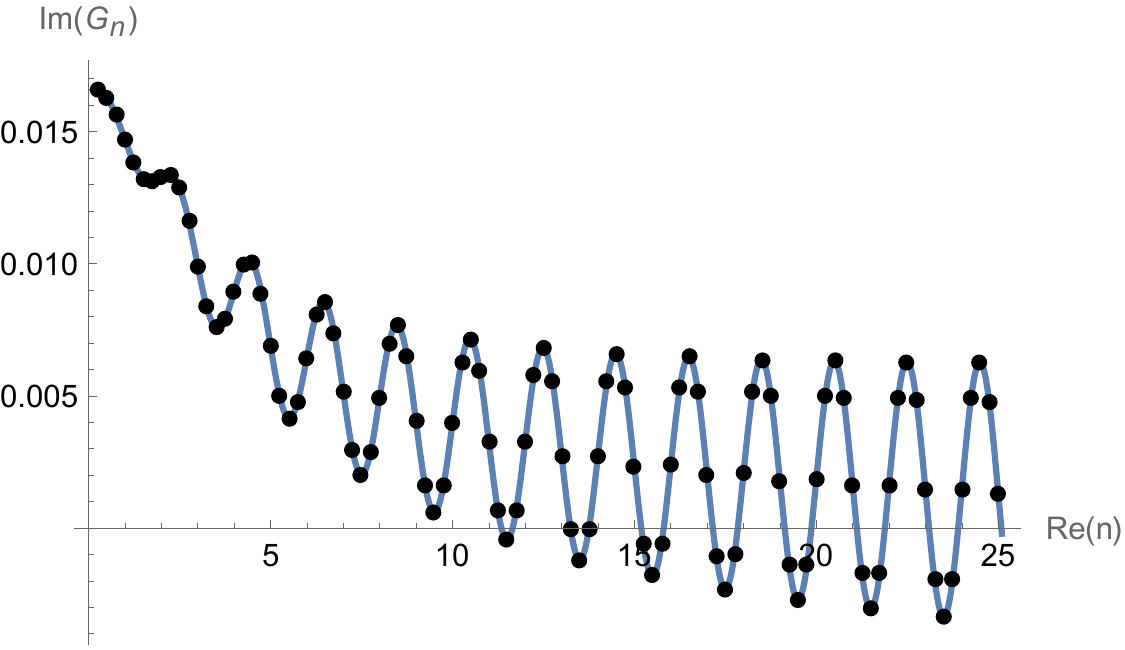}
		\caption{$\text{Im}(G_n)$ with $\text{Im}(n)=1$}
	\end{subfigure}	
	\begin{subfigure}{0.47\textwidth}
	\raggedright
		\includegraphics[width=1\linewidth]{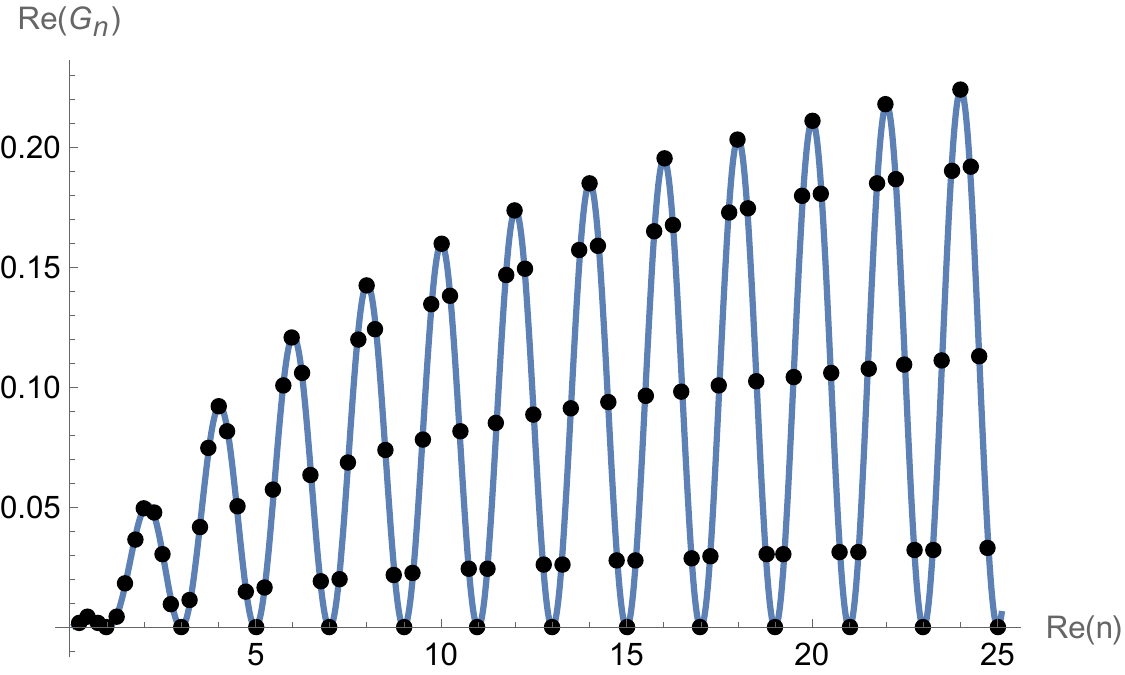}
		\caption{$\text{Re}(G_n)$ with $\text{Im}(n)=0$}
	\end{subfigure}	
	\qquad
	\begin{subfigure}{0.47\textwidth}
	\raggedright
		\includegraphics[width=1\linewidth]{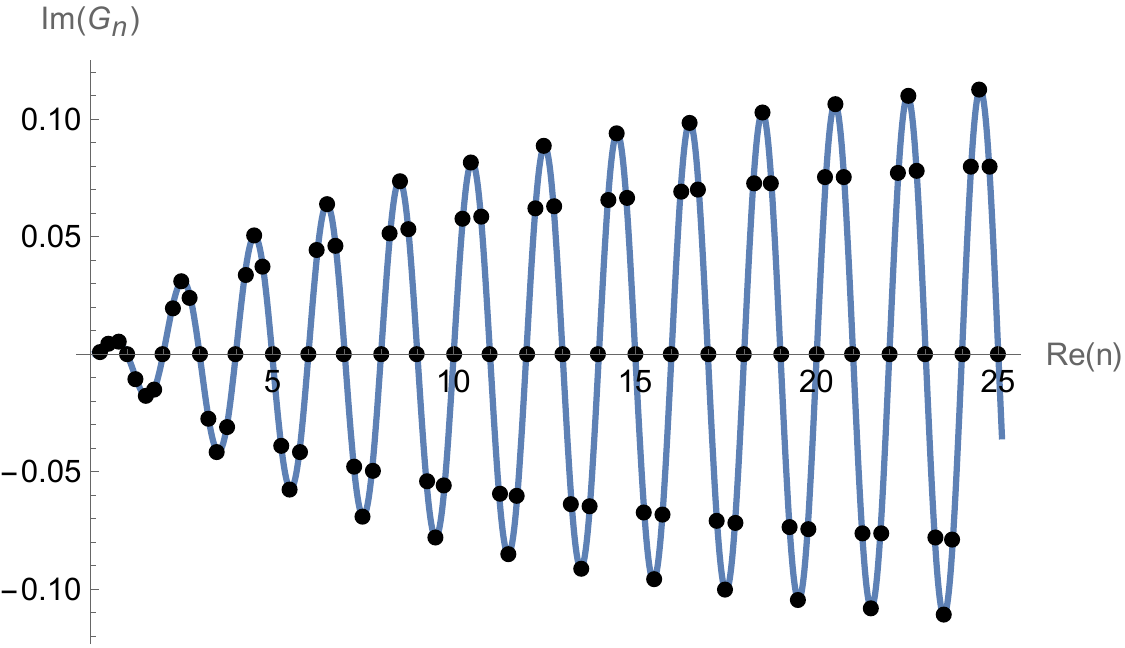}
		\caption{$\text{Im}(G_n)$ with $\text{Im}(n)=0$}
	\end{subfigure}	
	\begin{subfigure}{0.47\textwidth}
	\raggedright
		\includegraphics[width=1\linewidth]{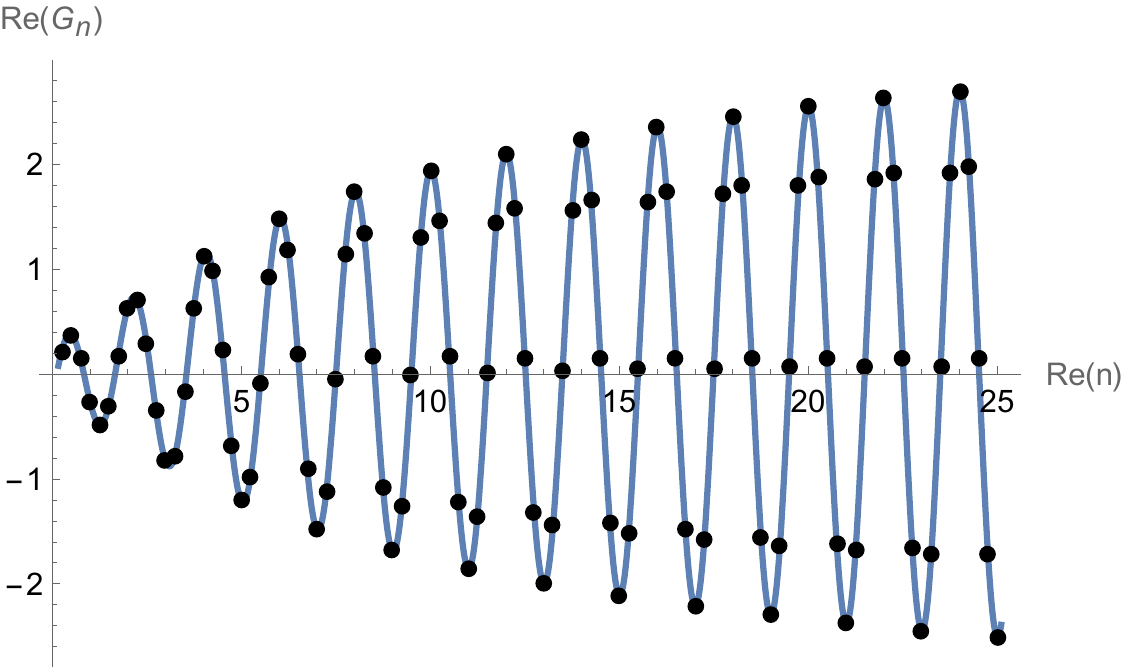}
		\caption{$\text{Re}(G_n)$ with $\text{Im}(n)=-1$}
	\end{subfigure}
	\qquad
	\begin{subfigure}{0.47\textwidth}
	\raggedright
		\includegraphics[width=1\linewidth]{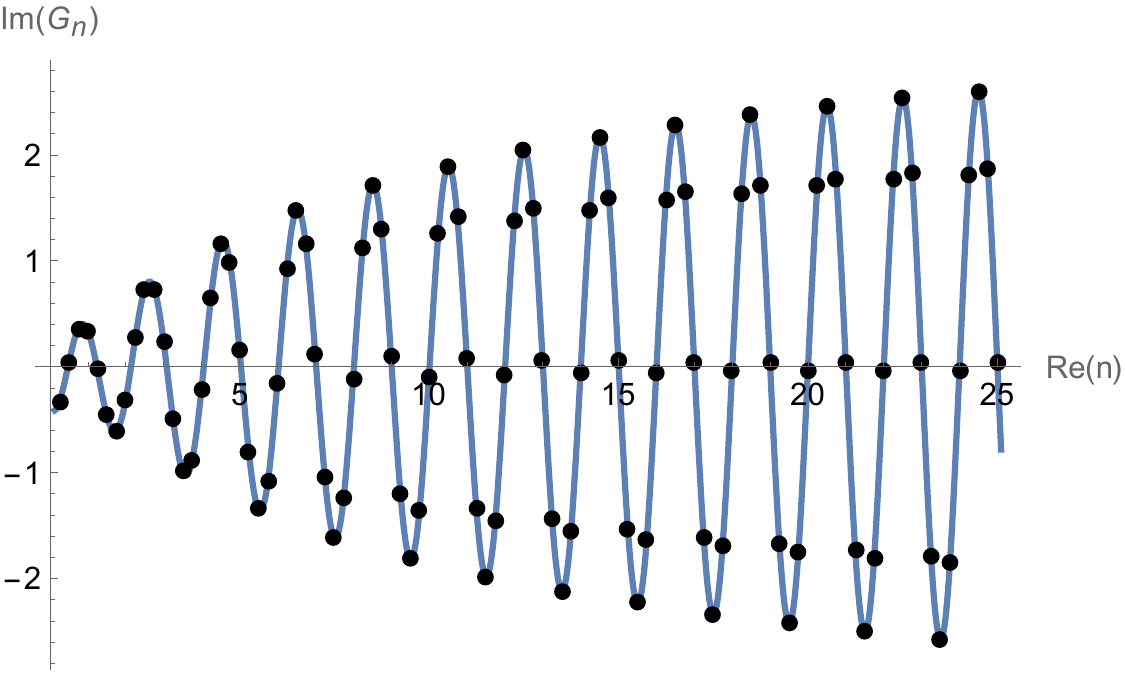}
		\caption{$\text{Im}(G_n)$ with $\text{Im}(n)=-1$}
	\end{subfigure}	
		\caption{
		Green's functions $G_n=\<\phi^n\>$ of the Hermitian quartic oscillator 
		$H=p^2+x^2+x^4$ as a function of the real part of $n$. 
	   	The imaginary part of $n$ is fixed to be $\text{Im}(n)=-1,0,1$. 
		The values of the black dots at $\text{Re}(n)=p/4$ with integer $p$ are computed 
		using the standard Hermitian inner product \eqref{Hermitian-inner-product} 
		and the ground-state wavefunction from the Hamiltonian diagonalization. 
		The blue curves are associated with the minimally singular solution 
		\eqref{quartic-large-n-expansion} with \eqref{quartic-a0}, \eqref{quartic-E0}. 
		The black dots are well interpolated by the blue curves. 
		The Green's functions have been divided by the leading large-$n$ asymptotic behavior, 
		as in the other figures for $G_n$. 
		}
	\label{fig_x4-trajectory-fixed-Im(n)}
\end{figure}

\begin{figure}
\begin{subfigure}{0.47\textwidth}
	\raggedright
		\includegraphics[width=1\linewidth]{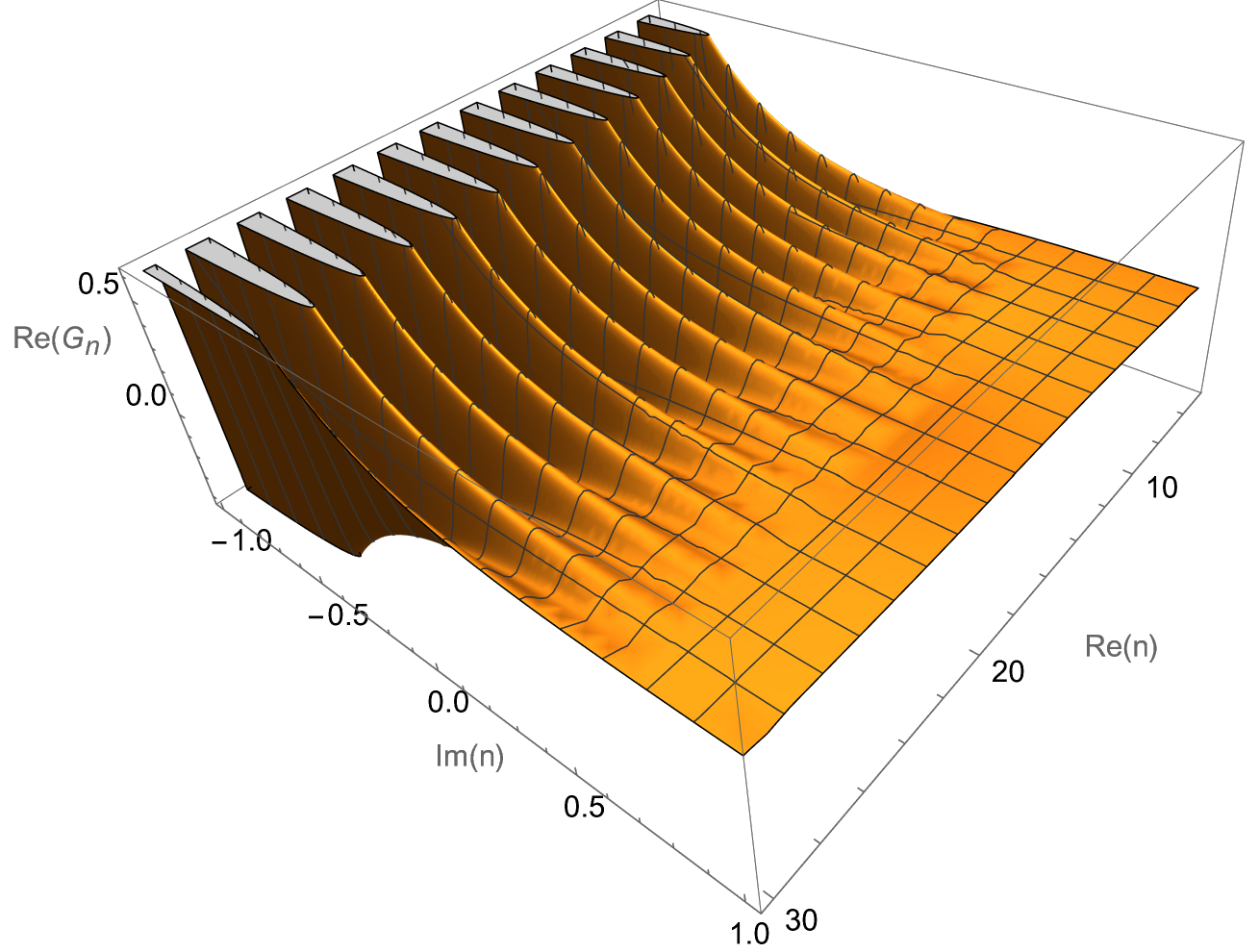}
		\caption{$\text{Re}(G_n)$}
	\end{subfigure}
	\qquad
	\begin{subfigure}{0.47\textwidth}
	\raggedright
		\includegraphics[width=1\linewidth]{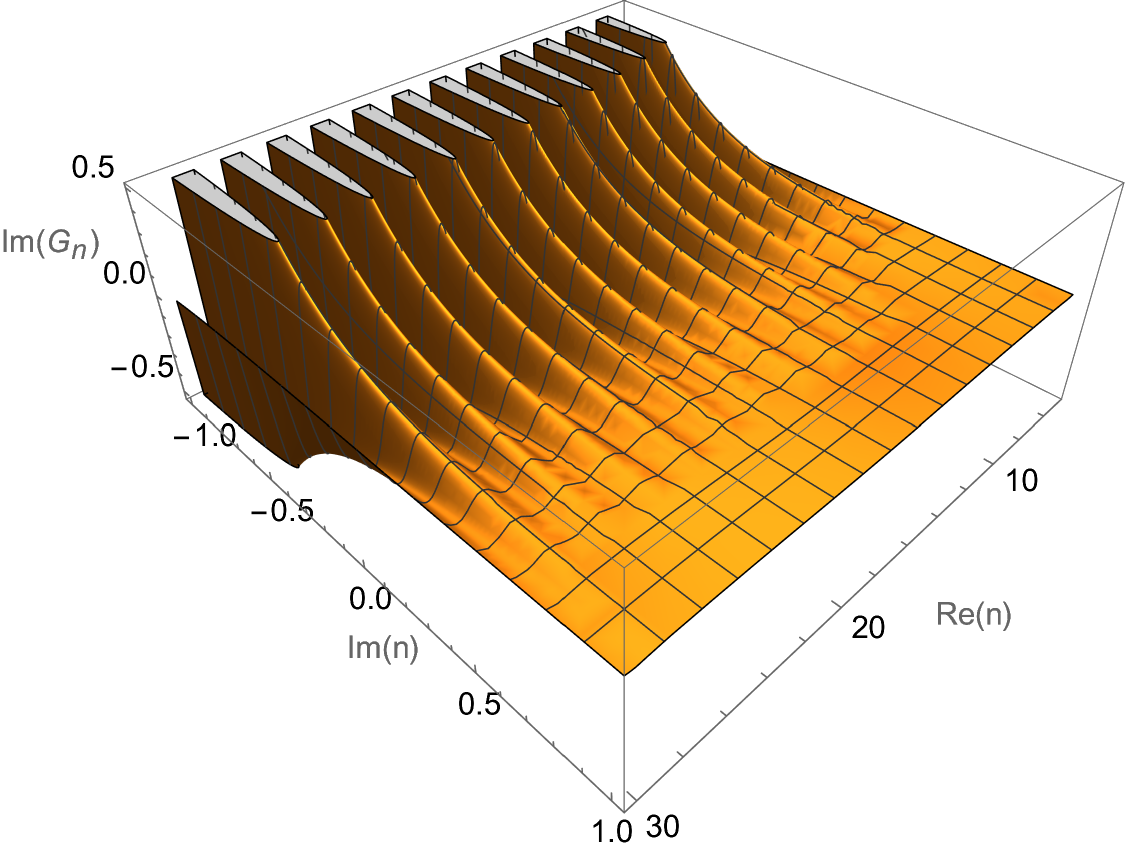}
		\caption{$\text{Im}(G_n)$}
	\end{subfigure}	
		\caption{
		Ground state Green's function $G_n=\<\phi^n\>$ of 
		the Hermitian quartic oscillator $H=p^2+x^2+x^4$ as a function of complex $n$. 
		They are computed by the large $n$ expansion of the minimally singular solution 
		\eqref{quartic-large-n-expansion} with \eqref{quartic-a0}, \eqref{quartic-E0}. The $1/n$ series is truncated to order $n^{-5}$. 
		}
	\label{fig_x4-trajectory-complex-n}
\end{figure}

Alternatively, we can study $G_n=\<\phi^n\>$ in the standard wave function formulation. 
In terms of the harmonic oscillator eigenfunctions, the matrix elements of the Hamiltonian \eqref{quartic-H} can be computed analytically. 
The diagonalization of a truncated Hamiltonian gives good approximations 
for the energy eigenvalues and eigenfunctions. 
We then use the approximate wave function $\psi[\phi]$ to compute $G_n$ 
based on the standard Hermitian inner product
\be
G_n=\<\phi^n\>=\int_{-\infty}^\infty d\phi\, \psi^\ast[\phi]\, \phi^n\,\psi[\phi]\,.
\label{Hermitian-inner-product}
\ee
The ground state estimates for $a_0$ and $G_2$ agree well with the $\phi^n$ trajectory bootstrap results in 
\eqref{quartic-E0} and \eqref{quartic-G2}. 
It is usually assumed that $n$ is a positive integer, 
but there is no obstruction to evaluating the integral in \eqref{Hermitian-inner-product} for complex $n$. 
In Fig. \ref{fig_x4-trajectory-fixed-Im(n)}, we compare the results from the wave function formulation and the minimally singular solution \eqref{quartic-large-n-expansion} with \eqref{quartic-a0}, \eqref{quartic-E0}. 
We find perfect agreement for both real and complex $n$. 
In Fig. \ref{fig_x4-trajectory-complex-n}, we further present the real and imaginary parts of $G_n$ as a function of complex $n$. 
For real $n$, the imaginary part of $G_n$ vanishes only at integral $n$, 
which is associated with the factor $1+e^{2\pi i\frac n 2}$ due to parity symmetry.

For relatively large $\text{Re}(n)$, the minimally singular solution can be evaluated directly using the $1/n$ series \eqref{quartic-large-n-expansion}. 
However, 
the direct evaluation at small $\text{Re}(n)$ is not accurate 
due to the asymptotic nature of the large $n$ expansion. 
To resolve this issue, 
we use the recursion relation \eqref{recursion-x2m} to express $G_n$ at small $\text{Re}(n)$ in terms of $G_n$ at relatively large $\text{Re}(n)$. 
In this way,  we can evaluate the minimally singular solution accurately at small $\text{Re}(n)$ as well. 
This is also the basic idea behind the matching procedure.

\begin{figure}
\begin{subfigure}{0.47\textwidth}
	\raggedright
		\includegraphics[width=1\linewidth]{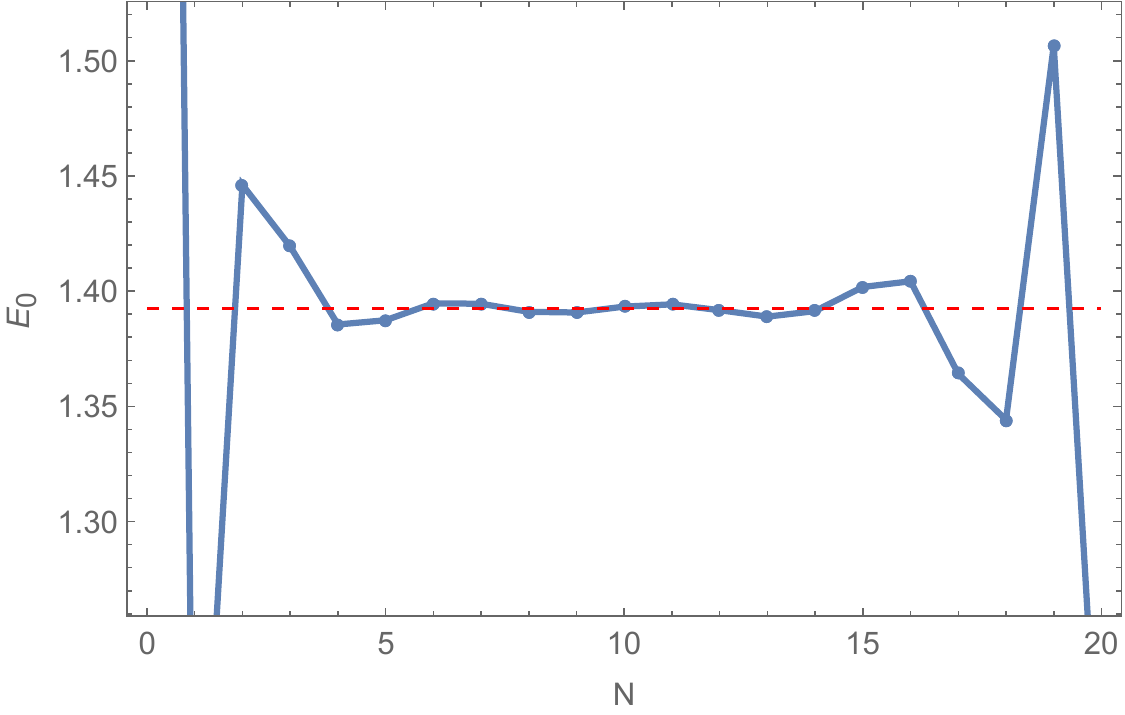}
		\caption{$M=8$}
	\end{subfigure}
	\qquad
	\begin{subfigure}{0.47\textwidth}
	\raggedright
		\includegraphics[width=1.02\linewidth]{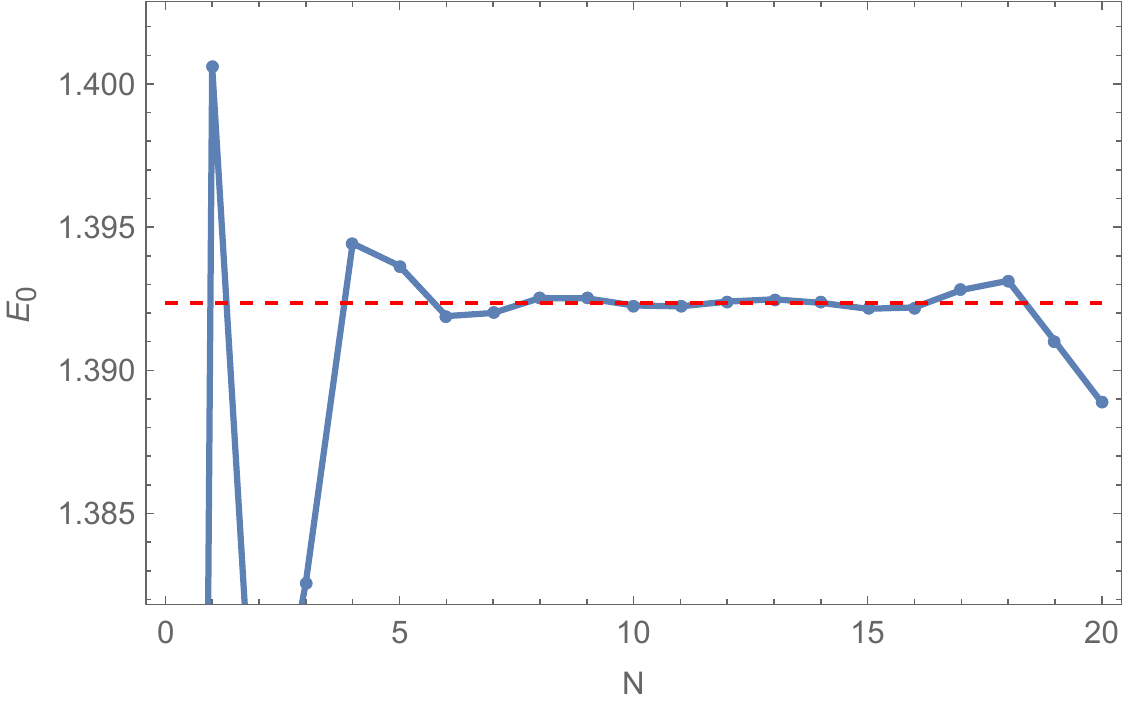}
		\caption{$M=10$}
	\end{subfigure}	
		\caption{Bootstrap results for the ground state energy of the Hermitian quartic oscillator $H=p^2+x^2+x^4$, where the matching order $M$ is fixed. We consider the concrete examples of $M=8$ and $M=10$, where $M$ determines the choice of matching conditions in \eqref{matching-x4}. As the $1/n$ series truncation order $N$ increases, the oscillatory estimates first approach the correct value and then exhibit growing deviations due to the asymptotic nature of the $1/n$ series. 
		In both cases, there is a plateau region around $N=M+2$, where the estimates are close to the dashed line (red) associated with the  accurate value of $E_0$ in \eqref{quartic-E0}. }
	\label{fig_x4-fixed-M}
\end{figure}

If the diagonalization method or other alternative methods are not available, we need to 
know how to extract the reasonable predictions from the bootstrap results, 
which depend on the matching order $M$ and the $1/n$ series truncation order $N$. 
The $1/n$ series is expected to be asymptotic, i.e., not a converging series.
\footnote{Although Stirling's formula does not lead to a converging power series for the gamma function $\G(n)$ in $1/n$, the combination $\ln \G(n)-(n\ln n-n+\frac 1 2 \ln \frac {2\pi}{n})$ admits a converging series expansion in terms of inverted rising factorials $1/(n+1)_j$. Can we extend this to the large $n$ expansion of a generic $G_n$? } 
For a fixed $M$, the results should first improve and then deteriorate as the series truncation order $N$ increases. 
In Fig. \ref{fig_x4-fixed-M}, we present the results for the ground state energy $E_0$ with $M=8,10$. 
In both cases, we find a plateau region around $N=M+2$. 
As expected, the good estimates for $E_0$ are around the center of the plateau. 
For $M=8$, the results for the plateau region $6\leq N\leq 14$ are
$(1.39473, 1.39454, 1.39104, 1.39074, 1.39331, 1.39429, 1.39174, 
1.38896, 1.39136)$. 
The mean value is about $1.3923$ and the standard deviation is about $0.002$, 
so the prediction is $E_{0,\text{mean}}^{(M=8)}=1.392(2)$, which is consistent with the diagonalization result.  
As the precise range of the plateau has some ambiguities, 
the mean value depends on our choice, 
but the standard deviation is of the same order and provides a proper error estimate. 
As in \cite{Li:2023nip}, a less ambiguous method is to iteratively discard 
the most distance solution from the average, 
so the good estimates are selected based on the distribution density. 
We obtain $(1.39074, 1.39104, 1.39136)$ and the corresponding prediction is $E^{(M=8)}_{0,\text{density}}=1.391(2)$. 
For $M=10$, the plateau range is about $8\leq N \leq 16$, 
then the mean value and standard deviation imply 
$E^{(M=10)}_{0,\text{mean}}=1.39235(14)$. 
Using the less ambiguous method, the last three numbers from iteratively discarding the most distance solution from the average are 
$(1.39248, 1.39251, 1.39252)$, so the resulting prediction is $E^{(M=10)}_{0,\text{density}}=1.39251(14)$,
which is more accurate than the $M=8$ prediction. 

\begin{figure}[h]
	\centering
		\includegraphics[width=0.7\linewidth]{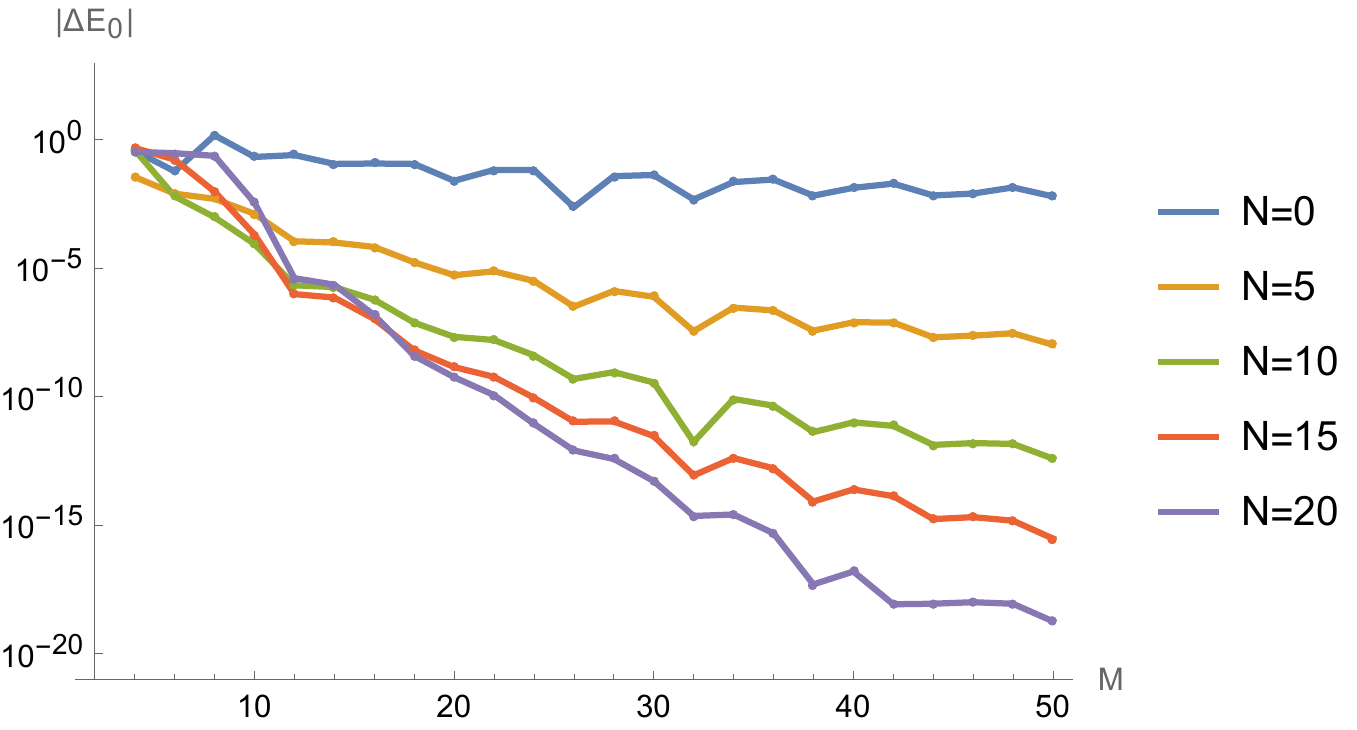}
	\caption{Absolute error in the ground state energy $E_0$ of the Hermitian quartic oscillator $H=p^2+x^2+x^4$, where the truncation order of the $1/n$ series is given by  $N=0,5,10,15,20$. 
	We use the accurate value for $E_0$ in \eqref{quartic-E0} as the reference value. 
	The bootstrap results are rapidly converging as the matching order $M$ increases. 
	At small $M$, the larger $N$ results could be worse due to the asymptotic nature of the large $n$ expansion. }
	\label{fig_x4-fixed-N}
\end{figure}

A truncated $1/n$ series provides more and more accurate approximations as $n$ increases.  
Therefore, a different approach is to consider a fixed series truncation order $N$ 
and examine the convergence properties 
as the matching order $M$ increases.  
As we solve the nonperturbative Green's functions exactly, the Green's functions at greater $n$ are as accurate as those at small $n$ if the initial conditions are exact. 
Since a truncated $1/n$ series improves as $n$ grows,  
the bootstrap solutions should converge to the correct values as the matching order $M$ increases. 
In Fig. \ref{fig_x4-fixed-N}, we show that the results are indeed rapidly converging for $N=0,5,10,15,20$ as the matching order $M$ grows. 
We could present the explicit digits, 
but the use of \eqref{quartic-E0} as a reference value makes the rapid convergence more clear. 
In fact, it is computationally more expensive to solve the $1/n$ expansion to high order than to solve the recursion relation \eqref{recursion-x2m} non-perturbatively to high order, 
so the matching order $M$ is set to be much bigger than the series truncation order $N$, such as $M=10N$. 
When $M\gg N$, the larger $N$ results are more accurate than those with smaller $N$. 
In this way, the results converge rapidly as the truncation parameters $(M,N)$ increase. 
The stable digits furnish the reliable prediction, while the error can be estimated from the varying digits. 

\subsection{Higher even powers}
Let us apply the bootstrap method to 
the cases of  high powers, i.e.,  $V(\phi)=\phi^2+\phi^m$ with $m=6,8,10,12,14,16$. 
Although the number of free parameters grows with $m$, 
we still obtain highly accurate results using the same matching procedure. 
We first study the large $n$ asymptotic behavior. 
At large $n$, the leading behavior is determined by the leading terms in \eqref{recursion-x2m}
\be
n^3 G_n\sim 4n G_{n+m+2}\quad (n\rightarrow \infty)\,.
\ee
The general form of the leading asymptotic behavior is given by
\be
G_n\sim \left(\frac{m}{2}+1\right)^{\frac{2n}{m+2}}\,\left[\G\left(\frac{n}{m+2}\right)\right]^2\,
\sum_{k=0}^{m+1}\,a_k\,e^{2\pi i\frac{kn}{m+2}}
\quad (n\rightarrow \infty)\,.
\label{large-n-leading-x2m}
\ee
The parity symmetry implies $G_n$ vanishes for odd $n$, so we have $a_k=a_{k+m/2+1}$. 
As in the quartic case, we consider the minimally singular solution with $a_0=a_{m/2+1}\neq 0$. 
\footnote{Other minimally singular solutions may also have interesting physical interpretations. } 
If we take into account the subleading terms in \eqref{recursion-x2m}, 
there exists an additional factor, whose general form is $n^{-\frac{m-6}{2(m+2)}}$. 
For $m\geq 8$, the large $n$ expansion of a parity-symmetric solution reads 
\be
G_n&\sim&
a_0\frac{1+e^{2\pi i\frac n 2}}{2}
\left(\frac{m}{2}+1\right)^{\frac{2n}{m+2}}\,\left[\G\left(\frac{n}{m+2}\right)\right]^2n^{-\frac{m-6}{2(m+2)}}
\left(1+\sum_{j=m/2-3}^{(m/2+1)N}\,c_j\left(\frac n 2 \right)^{-\frac{j}{m/2+1}}\right)
\,,
\nn
\label{large-n-expansion-x-2m}
\ee
where the $1/n$ series is truncated to order $n^{-N}$. 
The series coefficients can be computed order by order. 
If $m$ is an odd integer, the coefficients $c_j$ vanish for odd $j$. 
For large $m$, the low order coefficients take some general forms. 
\footnote{It might be interesting to study the large $m$ expansion. }
For example, the low order nonzero coefficients for $m\geq 16$ are
\be
c_{m/2-3}=\frac {2} {m-6}\,,\quad
c_{m/2-1}=-\frac{2E}{m-2}\,,\quad
c_{m/2+1}=\frac{m^2-20m-4}{16(m+2)}\,.
\label{x2m-c-low}
\ee
In fact, the expression of $c_{m/2-1}$ applies to $m\geq 12$, while that of $c_{m/2-3}$ is valid for $m\geq 8$. 
As $m$ decreases, the $1/n$ expansion of the recursion relation can have degenerate exponents at low order, so the concrete expressions of $c_j$ can be different from the general forms in \eqref{x2m-c-low}. 
For $m=6$, the additional factor $n^{-1/4}$ is different from the generic form 
due to the degeneracy in the exponents of the $1/n$ expansion. 
For $m=4$, the large $n$ expansion has two additional factors, i.e., an expected factor $n^{1/6}$ and 
a special factor $e^{-(n/2)^{1/3}}$, as shown in \eqref{quartic-large-n-expansion}.

In the matching procedure, 
the free parameters in \eqref{x-2m-free-parameters} and $a_0$ 
can be determined by the matching conditions
\be
G_n^{(\text{n.p.})}=G_n^{(\text{p.})}\,,\quad
n=M, M+2,\dots, M+m\,,
\label{matching-x-2m}
\ee
where $M$ indicates the matching order, $G_n^{(\text{n.p.})}$ is the non-perturbative solutions for $G_n$ from \eqref{recursion-x2m}, 
and $G_n^{(\text{p.})}$ is given by the $1/n$ series in \eqref{large-n-expansion-x-2m}.  
The number of matching constraints grows with $m$ as there are more free parameters. 
\footnote{In the wave function formulation, this is related to the higher order differential equations in the momentum representation.} 
They give rise to a system of  polynomial equations in the free parameters. 

\begin{figure}[h]
	\centering
	\begin{subfigure}{0.47\textwidth}
		\raggedright
		\includegraphics[width=1\linewidth]{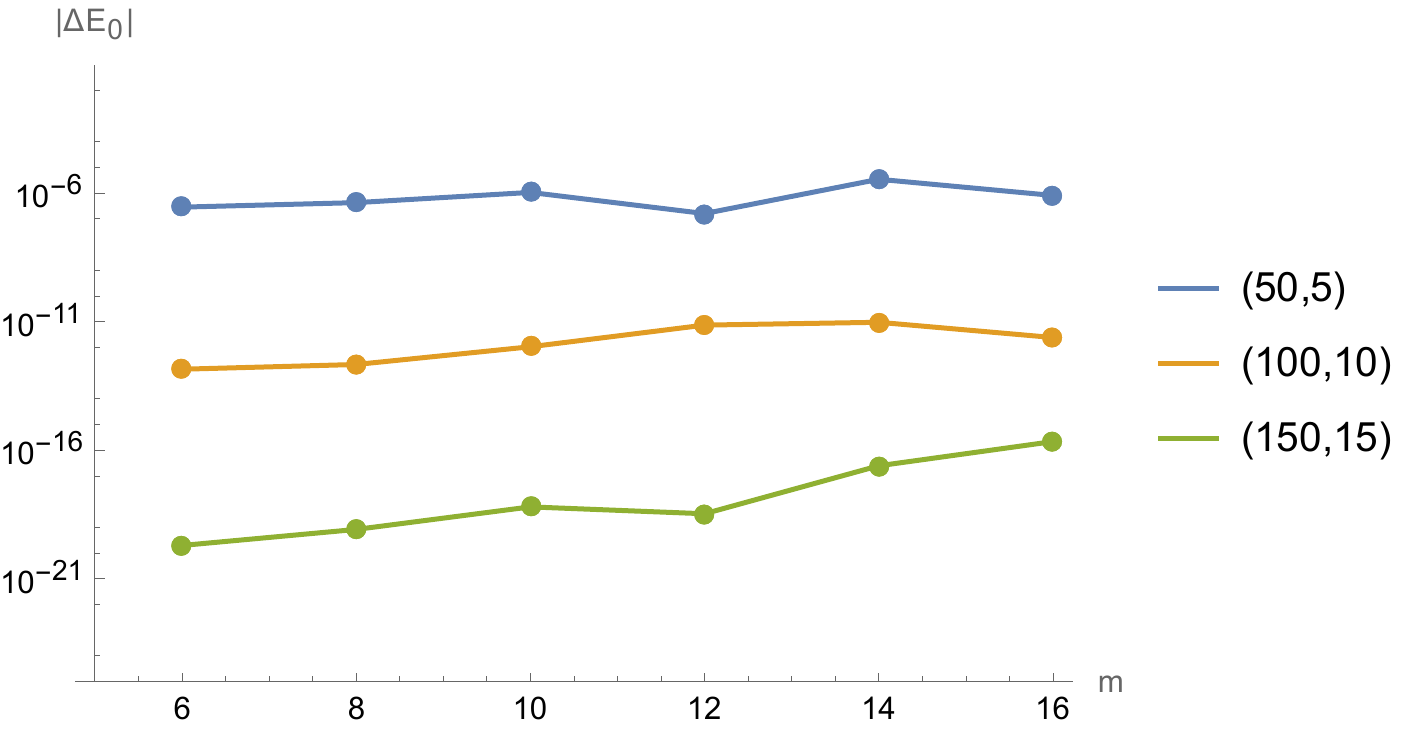}
		\caption{Matching $G_n$ with different $(M,N)$. }
	\end{subfigure}
	\quad
	\begin{subfigure}{0.47\textwidth}
		\raggedright
		\includegraphics[width=1\linewidth]{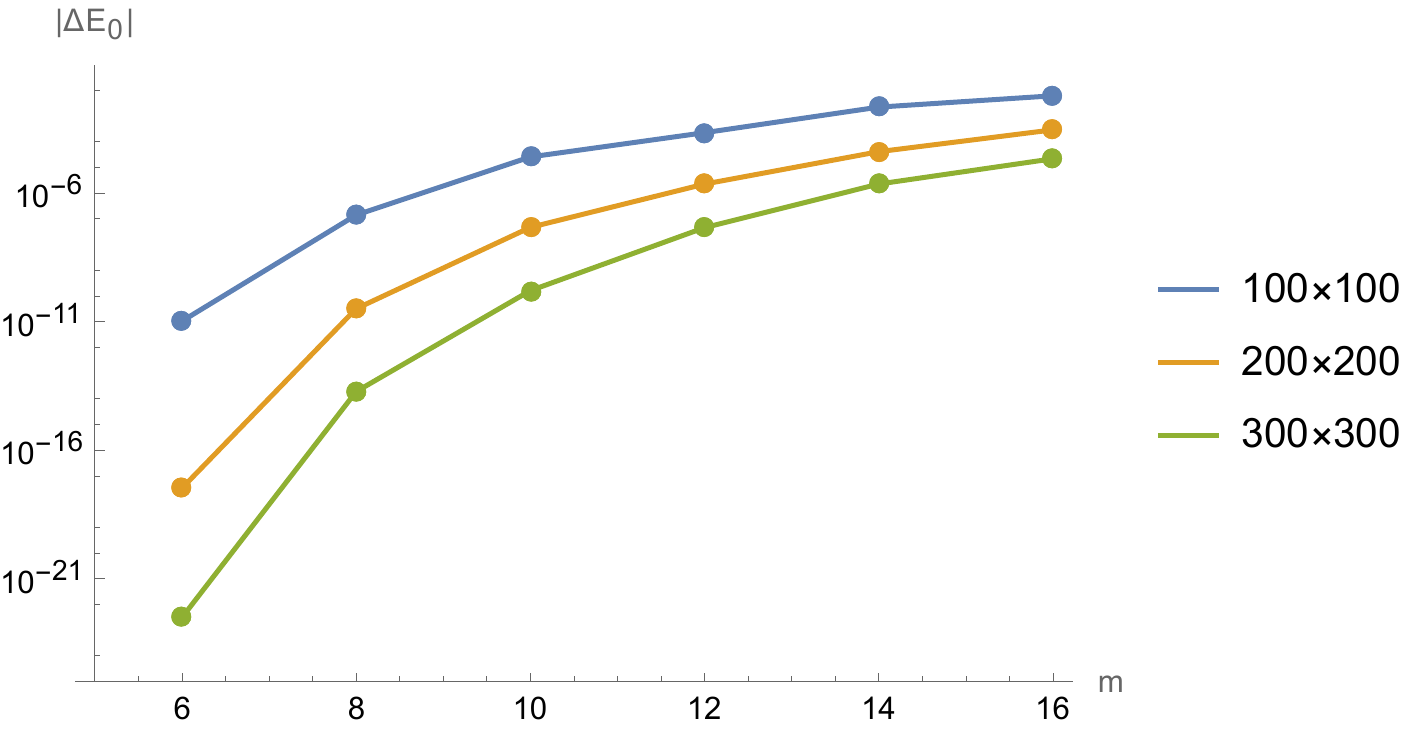}
		\caption{Diagonalizing $H$ with different sizes.}
	\end{subfigure}
	\caption{Absolute errors in the ground state energy $E_0$ for the anharmonic oscillator $H=p^2+\phi^2+\phi^m$ 
	from (a) the matching procedure of the $\phi^n$ trajectory bootstrap and 
	(b) the standard Hamiltonian diagonalization. 
	In our bootstrap method, 
	the matching conditions \eqref{matching-x-2m} are evaluated at $n=M, M+2, \dots, M+m$, 
	while the $1/n$ series \eqref{large-n-expansion-x-2m} is truncated to order $n^{-N}$. 
	The matrix elements of the truncated Hamiltonian are computed in the truncated bases of the harmonic oscillator eigenfunctions, which contain $100k$ low-lying eigenfunctions with $k=1,2,3$. 
	As $m$ grows, the errors in the diagonalization results increase significantly, 
	but our bootstrap method still gives highly accurate results. }
	\label{fig_x-2m}
\end{figure}

As $m$ increases, it becomes more and more challenging to solve 
a large set of high degree polynomial equations involving multiple variables. 
Note that these equations are linear in $G_2,\dots, G_{m-2}$, and $a_0$. 
Only the dependence on $E$ is nonlinear. 
For Hermitian solutions, the energy are usually expected to be real and the complex solutions are irrelevant. 
As in the case of the harmonic oscillator example, we introduce the $\eta$ function \eqref{eta-HO} 
and transform the difficult problem of solving a large set of high degree polynomial equations 
into an easier minimization problem. 
After deriving the explicit expressions of $G_n^{(\text{n.p.})}$ and $G_n^{(\text{p.})}$, 
we scan the real $E$ and search for the local minima of $\eta$ with $\eta_\text{min}=0$, 
which leads to highly accurate results for the energies and expectation values.

In Fig. \ref{fig_x-2m}, we compare the errors in the ground state energy $E_0$ from 
the matching approach and the Hamiltonian diagonalization approach for $m=6,8,\dots,16$. 
To remind the reader, the parity invariant potential is given by $V(\phi)=\phi^2+\phi^m$. 
As the power $m$ increases, the accuracy of the diagonalization results decreases rapidly, 
but the matching conditions still lead to highly accurate results for 
the same set of truncation orders $(M,N)$. 
Our bootstrap method appears to be more efficient, especially at larger $m$. 

As we do not use any positivity constraints, 
the matching procedure also applies to the non-Hermitian models, 
which can violate some positivity assumptions. 
Below, we extend the discussion to the non-Hermitian $\mathcal {PT}$ invariant oscillators.

\section{The $\mathcal{PT}$ invariant potential $V(\phi)=-(i\phi)^{m}$}
\label{sec:ixm}
In this section, we consider the non-Hermitian Hamiltonian
\be
H=p^2-(i\phi)^m\,,
\label{H-ixm}
\ee
which is invariant under the $\mathcal {PT}$ transformation
\be
\phi\rightarrow -\phi\,,\quad
i\rightarrow -i\,.
\ee
In accordance with $\mathcal {PT}$ symmetry, we define the Green's functions as
\be
G_n=\<(i\phi)^n\>\,.
\ee 
Note that $i\phi$ plays an analogous role as $\phi^2$ in the parity invariant cases, 
but there is no simple vanishing constraint on $G_n$ associated with $\mathcal {PT}$ symmetry. 
\footnote{In the $\mathcal {PT}$ invariant theories, the analytically continued $G_n$ also vanishes at certain real $n$, but the vanishing points do not have equal spacing. 
It may also be interesting to consider the $m$ generalization of parity invariant theories, 
such as $(\phi^2)^{m/2}$ where $m$ is not an even integer. } 
The recursion relation \eqref{xn-recursion} becomes
\be
(n+1)_3G_n-4E(n+3)G_{n+2}=2(2n+m+6)G_{n+m+2}\,,
\label{ixm-recurion}
\ee
where the normalization is set by $G_0=1$. 
Note that $m$ is not necessarily an integer. 
We will discuss the case of fractional $m$ in Sec. \ref{sec:fractional-m} 
and the case of irrational $m$ in Sec. \ref{sec: irrational}. 
We will focus on some $\mathcal {PT}$ symmetric ground state solutions 
associated with $\mathcal {PT}$ symmetric quantization schemes.  
Their complete energy spectra are real and bounded from below. 

\subsection{Integral powers}
\begin{figure}[h]
	\begin{subfigure}{0.41\textwidth}
		\raggedright
		\includegraphics[width=1\linewidth]{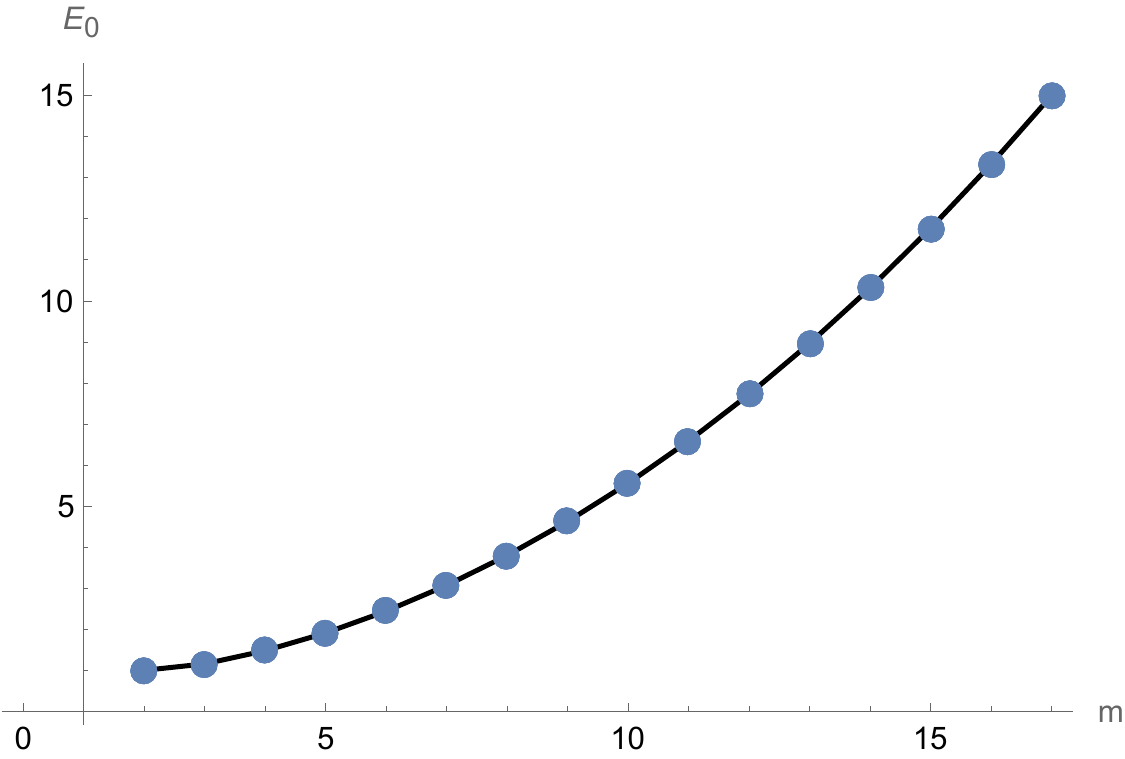}
		\caption{$2\leq m\leq 17$}
		\label{fig_ix-m-E0-a}
	\end{subfigure}
	\quad
	\begin{subfigure}{0.56\textwidth}
		\raggedright
		\includegraphics[width=1\linewidth]{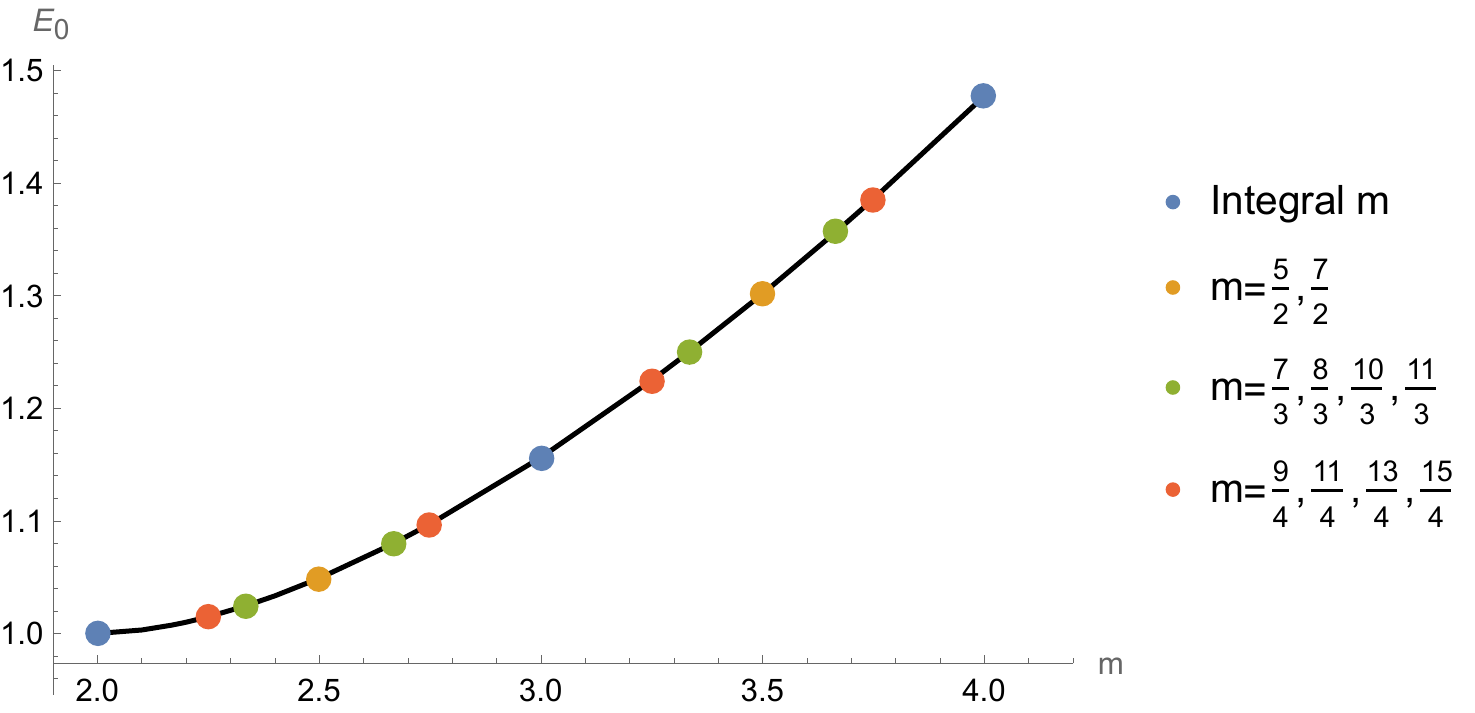}
		\caption{$2\leq m\leq 4$}
		\label{fig_ix-m-E0-b}
	\end{subfigure}
	\caption{Ground state energy $E_0$ of the non-Hermitian Hamiltonian $H=p^2-(i\phi)^m$ at various $m$. We focus on the $\mathcal {PT}$ symmetric cases. 
	In (a), we present the integral $m$ results for $2\leq m\leq 17$. 
	In (b), we zoom in on the range $2\leq m\leq 4$ and present some fractional $m$ results as well. }
	\label{fig_ix-m-E0}
\end{figure}

\begin{figure}[h]
	\begin{subfigure}{0.47\textwidth}
	\raggedright
		\includegraphics[width=1\linewidth]{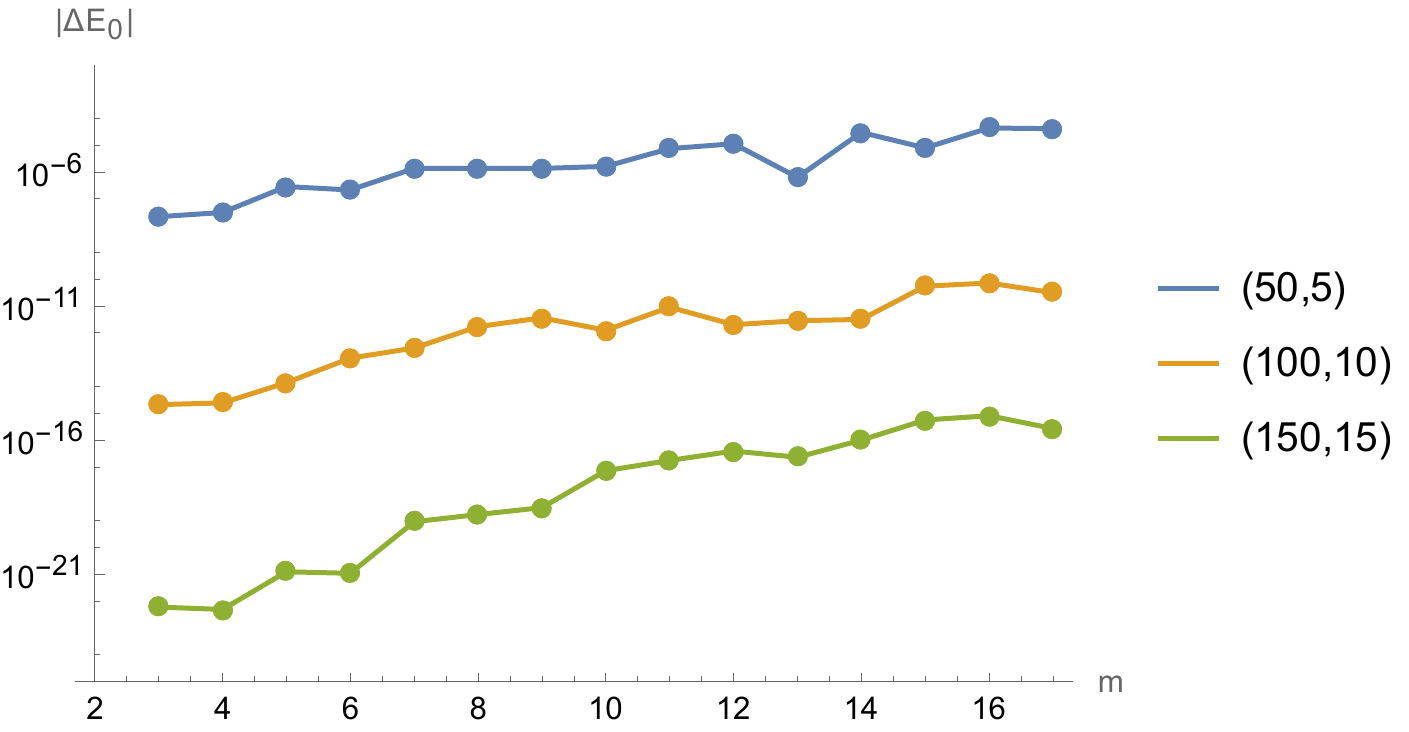}
		\caption{$m=3,4,\dots,17$}
		\label{fig_ix-m-error-a}
	\end{subfigure}
	\qquad
	\begin{subfigure}{0.47\textwidth}
	\raggedright
		\includegraphics[width=1\linewidth]{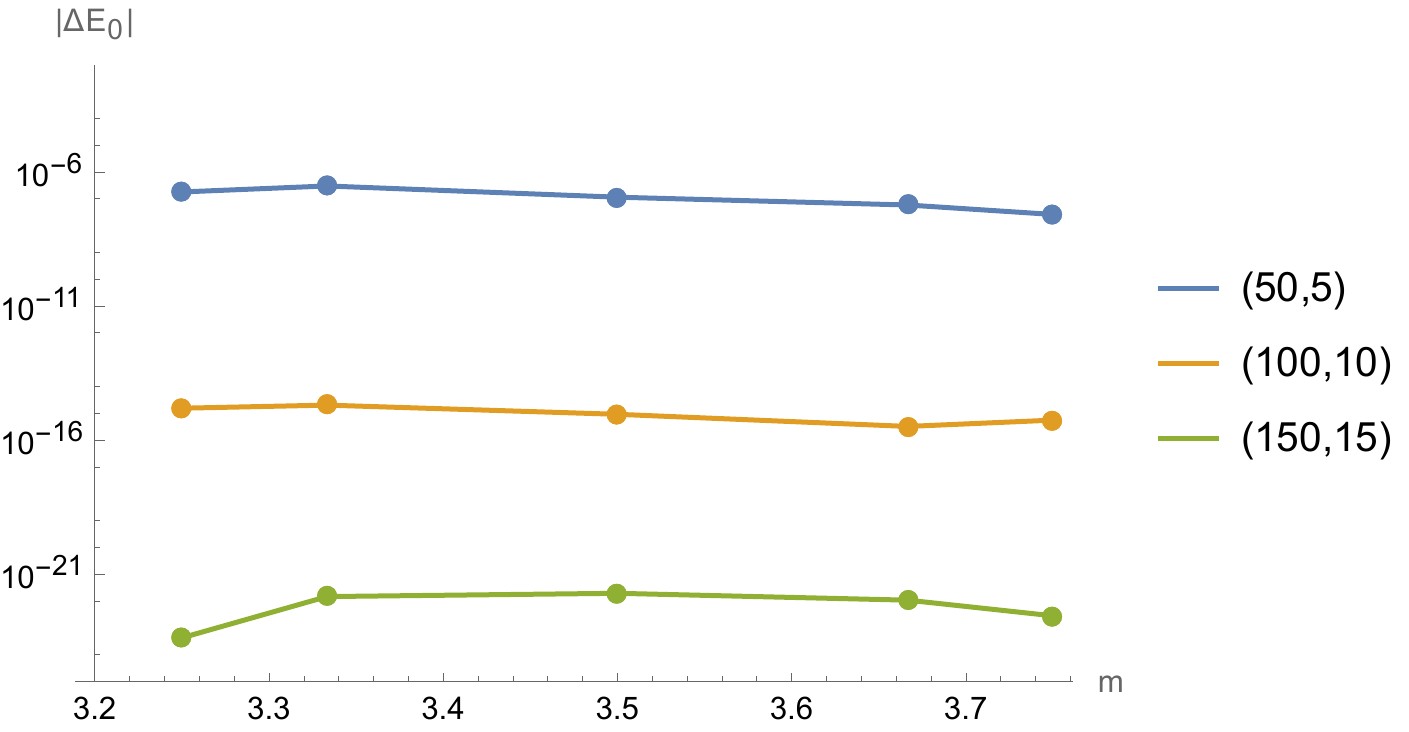}
		\caption{$m=\frac{13}{4},\frac{10}{3},\frac{7}{2},\frac{11}{3},\frac{15}{4}$}
		\label{fig_ix-m-error-b}
	\end{subfigure}	
		\caption{Absolute errors in the ground state energies of $H=p^2-(i\phi)^m$ 
		at various $m$ with $(M,N)=(50,5), (100,10), (150,15)$. We use $M$ to denote the matching order 
		and $N$ to indicate the truncation order of the $1/n$ series. 
		We present the integral $m$ results for $3\leq m\leq 17$.  
		Then we zoom in on the range $3\leq m\leq 4$ and present the results at some fractional $m$. }
	\label{fig_ix-m-error}
\end{figure}

Let us assume that $m\geq 3$ and $m$ is an integer. 
The independent set of free parameters is chosen to be
\be
(E\,,G_1\,,G_2\,,\dots,G_{m-2})\,.
\ee
The leading asymptotic behavior can be derived from 
\be
n^3G_n\sim 4n\,G_{n+m+2}\quad (n\rightarrow\infty), 
\ee
so the general leading behavior takes the same form as \eqref{large-n-leading-x2m}
\be
G_n\sim \left(\frac{m}{2}+1\right)^{\frac{2n}{m+2}}\,\left[\G\left(\frac{n}{m+2}\right)\right]^2\,
\sum_{k=0,\pm 1,\pm 2,\dots}\,a_k\,e^{2\pi i\frac{kn}{m+2}}
\quad (n\rightarrow \infty)\,.
\ee
After taking into account the subleading terms, we obtain the $1/n$ series
\be
G_n&\sim&\left(\frac m 2+1\right)^{\frac{2n}{m+2}}\left[\G\left(\frac n {m+2}\right)\right]^2
n^{-\frac{m-6}{2(m+2)}}
\sum_{k=0,\pm 1,\pm 2,\dots} a_k\,e^{2\pi i\frac{kn}{m+2}}
\left(1+\sum_{j=m-2}^{(m+2)N} 
c_{k,j}\left(\frac n 2\right)^{-\frac{j}{m+2}}\right)\,
\quad (n\rightarrow \infty)\,,\quad
\label{large-n-ixm}
\ee
which is truncated to order $n^{-N}$. 
For integral $n$, the number of independent $1/n$ series is $m+2$ and we can set infinitely many prefactors $a_k$ to zero. 
The coefficients $c_{k,j}=c_{k,j}[E]$ are functions of the energy $E$. 
Since the recursion relation \eqref{ixm-recurion} is invariant under the rotation
\footnote{The Hermitian case \eqref{recursion-x2m} also has some discrete rotation symmetry, as the coefficients of the mass and quartic term transform properly. 
One can also consider the rotation of $\hbar$. }
\be
G_n\rightarrow G_n\,e^{2\pi i\frac{kn}{m+2}}\,,\quad
E\rightarrow E\,e^{2\pi i\frac{(-2)k}{m+2}}\,,
\label{PT-rotation-symmetry-integral}
\ee
the $k\neq 0$ series coefficients are related to $c_{0,j}$ by
\be
c_{k,j}[E]=c_{0,j}\big[Ee^{2\pi i\frac{2k}{m+2}}\big]\,,
\ee
which generalizes the relation \eqref{c0c1-E} in the harmonic case $m=2$. 
For odd $m$, we also notice that $c_{k,j}[E]=c_{0,j}[E]\, e^{2\pi i\frac{(m+1)k}{2(m+2)}}$. 
At large $m$, the low order coefficients take the general forms
\be
c_{0,m-2}=\frac {2E}{m-2}\,,\quad
c_{0,m+2}=\frac{m^2-20m-4}{16(m+2)}\,,
\label{coefficient-1}
\ee
\be
c_{0,2m-4}=\frac{2E^2}{(m-2)^2}\,,\quad
c_{0,2m}=\frac{(5m^2-44m+28)E}{8(m-2)(m+2)}\,.
\label{coefficient-2}
\ee
We assume that the $\mathcal {PT}$ symmetry is not broken,  
which implies that $G_n=\<(i\phi)^n\>$ should be real for real $n$. 
Therefore, we have
\be
a_{-k}=(a_{k})^\ast\,.
\ee
We will focus on the $\mathcal {PT}$ symmetric solutions with only two nonvanishing prefactors $a_1$ and $a_{-1}$, 
which are minimally singular as there are only two asymptotic behaviors.  
\footnote{ 
The number of $\mathcal {PT}$ symmetric solutions grows with $m$. 
See \cite{Bender:2022eze,Bender:2023ttu} for more details about the $D=0$ case. 
For $m> 4$, certain choices of $k$ are related to wave functions that vanish as $\phi\rightarrow \pm \infty$, 
which is consistent with the results from the naive diagonalization of a truncated Hamiltonian. }
To determine the free parameters, we impose the matching conditions
\be
G_n^{(\text{n.p.})}=G_n^{(\text{p.})}\,,\quad
n=M,M+1,\dots,M+m\,.
\label{matching-ixm}
\ee
As before, $G_n^{(\text{n.p.})}$ are obtained by solving the recursion relation \eqref{ixm-recurion}. 
Their analytic expressions are at most linear in $(G_1\,,G_2\,,\dots,G_{m-2})$, 
but they can be of high degree in $E$. 
We again introduce the $\eta$ function \eqref{eta-HO}.
The solutions to the matching conditions \eqref{matching-ixm} 
correspond to the local minima with $\text{min}(\eta)=0$. 
In Fig. \ref{fig_ix-m-E0-a}, we present the results for the ground state energy $E_0$ at various integer $m$. 
In Fig. \ref{fig_ix-m-error-a}, we show that the absolute error in $E_0$ only grows mildly as $m$ increases. 

According to the accurate solutions from the matching procedure, 
we conjecture that the argument of $a_1$ takes a simple analytic form
\be
\arg[a_1]=\frac {m+10}{2(m+2)}\pi\,,
\label{PT-ratio}
\ee
which should be related to the choice of the Stokes sectors. 
Note that $a_1$ is negative real at $m=6$, as $\arg[a_1]=\pi$.  
The deviation of the numerical solution for $\arg[a_1]$ from the analytic expression \eqref{PT-ratio} also 
provides an error estimation for the bootstrap results. 

\begin{figure}[h]
	\begin{subfigure}{0.5\textwidth}
	\raggedright
		\includegraphics[width=1\linewidth]{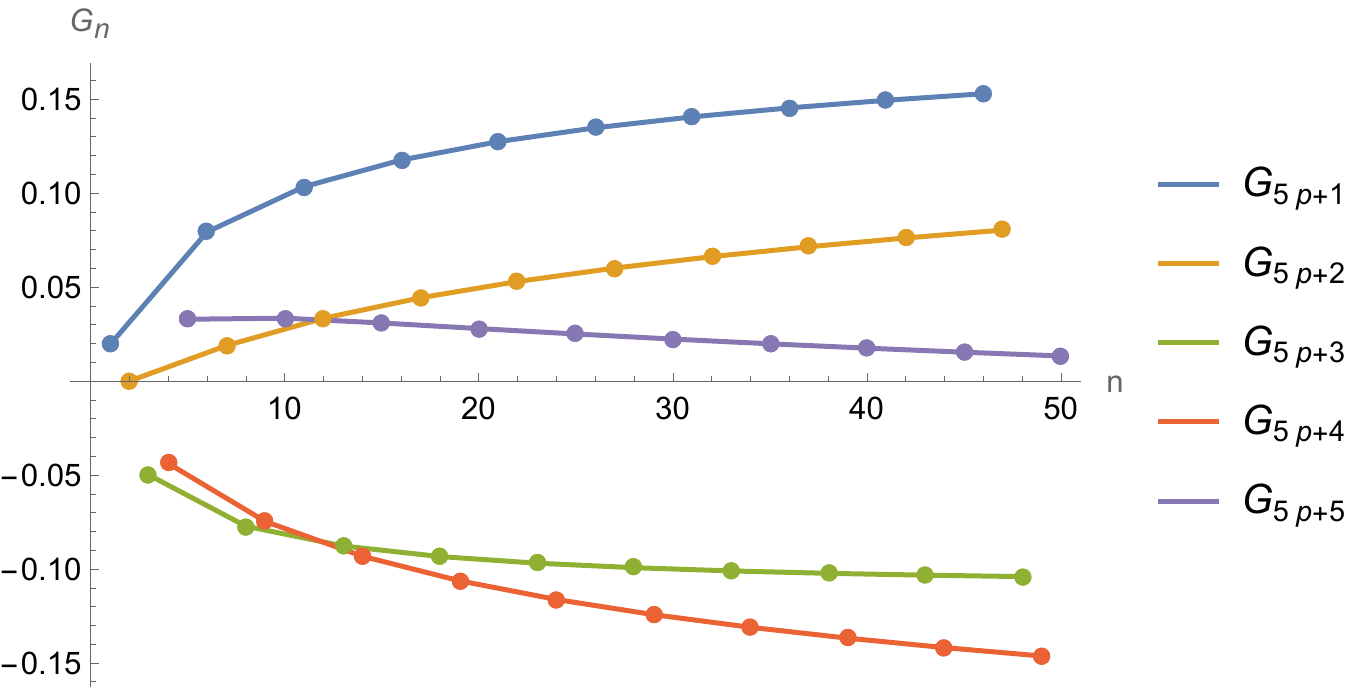}
		\caption{$n=1,2,\dots,50$}
		\label{fig_ix-3-a}
	\end{subfigure}
	\qquad
	\begin{subfigure}{0.45\textwidth}
	\raggedright
		\includegraphics[width=1\linewidth]{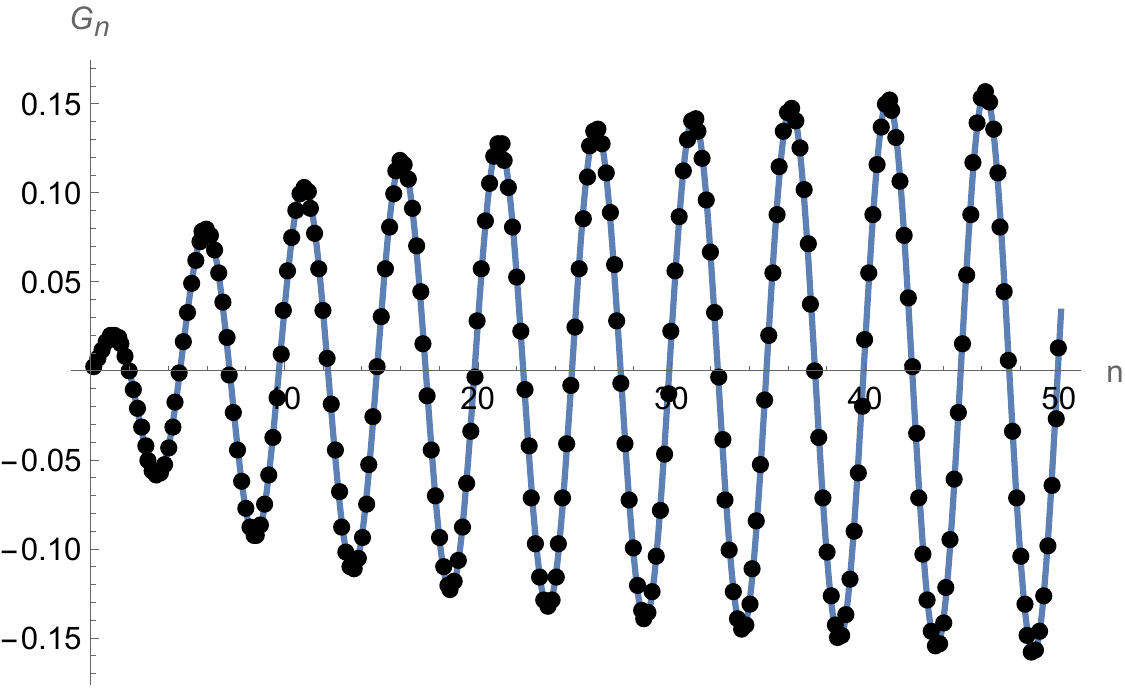}
		\caption{$n=\frac 1 5,\frac 2 5,\dots,\frac{249}{5},\frac{250}{5}$}
		\label{fig_ix-3-b}
	\end{subfigure}	
		\caption{Ground state Green's functions $G_n=\<(i\phi)^n\>$ of the non-Hermitian cubic theory $H=p^2-(i\phi)^3$. According to the integral $n$ results from the wave function formulation and the Hamiltonian diagonalization, one may naively find 5 branches of Green's functions, as indicated in (a). 
		However, as we also present the fractional $n$ results in (b),
		an oscillatory blue curve emerges, which is precisely given by the minimally singular solution with \eqref{sol-m-3-a1} and \eqref{sol-m-3-E}. 
		}
	\label{fig_ix-3}
\end{figure}

Before considering the cases with fractional power $m$, 
let us revisit the basic example of the cubic oscillator, i.e., $m=3$. 
As discussed in \cite{Li:2023ewe}, 
if we focus on $G_n$ with integer $n$, 
there seem to be 5 branches of Green's functions, 
corresponding to $G_{5p+k}$ with $k=1,2,3,4,5$, as shown in Fig. \ref{fig_ix-3-a}. 
However, the analytic continuation in $n$ allows us to consider $G_n$ at non-integer $n$. 
One may wonder if the non-integer $n$ cases correspond to more exotic branches of solutions.  

As in the Hermitian cases in Sec. \ref{sec:x2m}, 
we can also study the non-Hermitian $G_n$ using the wave function formulation. 
According to the symmetry of the Hamiltonian \eqref{H-ixm}, 
it is natural to use the $\mathcal {PT}$ inner product
\be
G_n=\<(i\phi)^n\>=\int_{-\infty}^\infty d\phi\, \{\psi[-\phi]\}^\ast\, (i\phi)^n\,\psi[\phi]\,,
\label{PT-inner-product}
\ee
which is valid for $1<m<4$. 
\footnote{In this range, the Stokes sectors contain the real axis of $\phi$ and 
at least one eigenvalue of \eqref{H-ixm} is real. } 
In contrast to the Hermitian inner product, 
the $\mathcal {PT}$ inner product does not obey positivity constraints. 
We assume that the wave function $\psi[\phi]$ is $\mathcal {PT}$ symmetric. 
In Fig. \ref{fig_ix-3-b}, we present the ground-state results from the wave function formulation at fractional $n=p/5$, where $p$ is a positive integer. 
It is clear that they are interpolated by an oscillatory curve, 
which is similar to the quartic case in Fig. \ref{fig_x4-trajectory-fixed-Im(n)}. 
The interpolating function is precisely a minimally singular solution for the ground state.  
The large $n$ asymptotic behaviors of $\<(i\phi)^n\>$ are associated with 
the two nonvanishing prefactors $a_{1}$ and $ a_{-1}=(a_1)^\ast$:
\be
a_1&=&-0.128537084089570612897940053524...\nn
&&-0.176916118650967924857364104948...i\,,
\label{sol-m-3-a1}
\\
E&=&1.15626707198811329379921917800...\,,
\label{sol-m-3-E}
\\
G_1&=&0.590072533090700847855025174549...\,,
\ee
where the argument of $a_1$ is consistent with the analytic expression \eqref{PT-ratio}. 
Therefore, the 5 naively different branches of Green's functions at integer $n$ are unified by the oscillatory minimally singular solution. 
There is only one interpolating solution for both the integer and non-integer $n$. 
The unification by analytic continuation is one of the powerful aspects of analyticity, 
which is not restricted to the cubic example. 
As in the quartic case, we again use the nonperturbative recursion relation \eqref{ixm-recurion} to derive accurate results for $G_n$ at small $\text{Re}(n)$.

\subsection{Fractional powers}
\label{sec:fractional-m}
In the Hermitian quartic and non-Hermitian cubic cases, 
we verify that the solutions for $G_n$ at non-integer $n$ are consistent with the results from the more standard wave function formulation, 
so the $n$ complexification is not a purely mathematical trick. 
In fact, if we want to bootstrap the $\mathcal {PT}$ invariant cases with non-integer power $m$, 
it is inevitable to consider $\<(i\phi)^n\>$ with non-integer $n$,  
which was one of the main motivations for considering the analytical continuation in $n$ \cite{Li:2022prn}. 

For simplicity, we will focus on the recursion relation \eqref{ixm-recurion} with fractional $m$, 
but the general procedure can be extended to irrational $m$, which will be demonstrated later in Sec. \ref{sec: irrational}. 
In order to compare with the results from the wave function formulation, 
we restrict the range of $m$ to $2<m<4$ where the standard diagonalization method is valid. 
In Fig. \ref{fig_ix-m-E0-b}, we present some results for the ground state energy at fractional $m$, 
which leads to a smooth interpolating curve. 
They are also consistent with the Runge-Kutta result in the classical work on $\mathcal {PT}$ invariant quantum mechanics  \cite{Bender:1998ke}.  
In Fig. \ref{fig_ix-m-error-b}, we show that the results for the ground state energy remain highly accurate for fractional $m$, as in the integer $m$ cases. 

\begin{figure}[h]
	\begin{subfigure}{0.47\textwidth}
	\raggedright
		\includegraphics[width=1\linewidth]{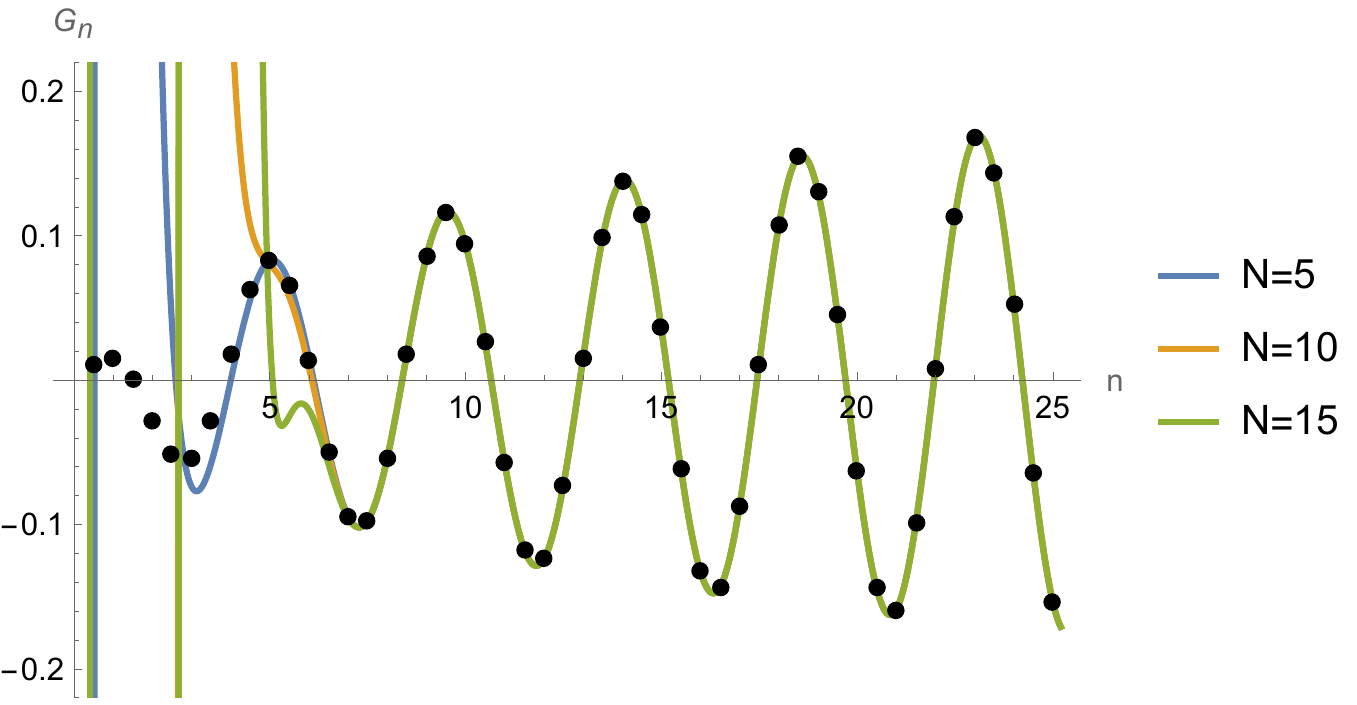}
		\caption{$m=\frac{5}{2},\,n=\frac {p}{2}$}
		\label{subfig:trajectory-5-2}
	\end{subfigure}
	\qquad
	\begin{subfigure}{0.47\textwidth}
	\raggedright
		\includegraphics[width=1\linewidth]{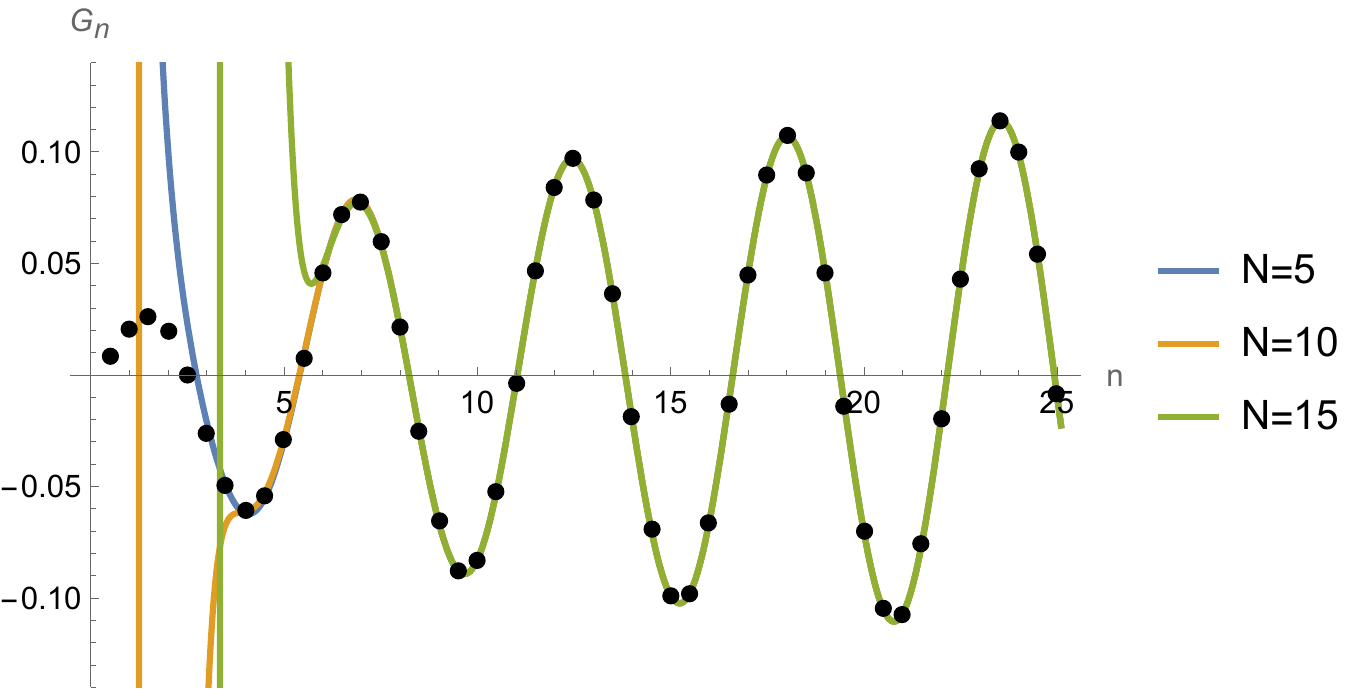}
		\caption{$m=\frac{7}{2},\,n=\frac {p}{2}$}
	\end{subfigure}	
	\begin{subfigure}{0.47\textwidth}
	\raggedright
		\includegraphics[width=1\linewidth]{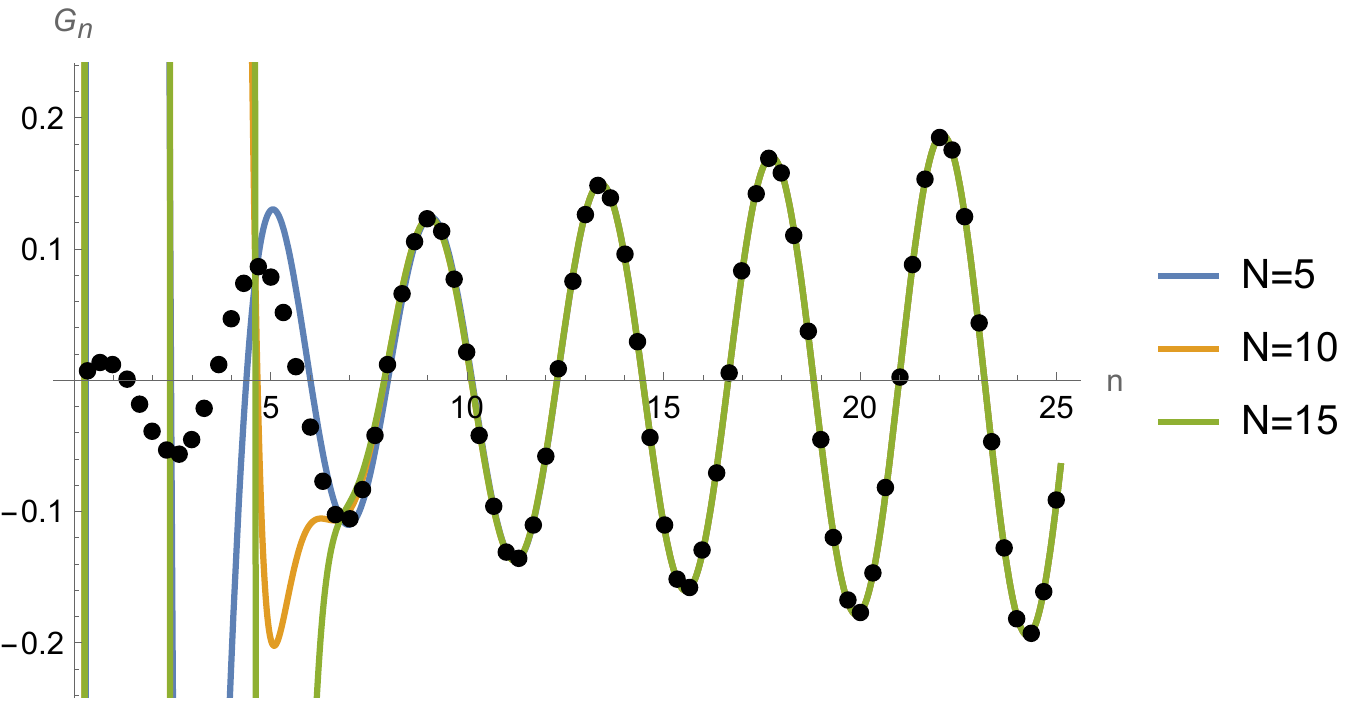}
		\caption{$m=\frac{7}{3},\,n=\frac {p}{3}$}
	\end{subfigure}
	\qquad
	\begin{subfigure}{0.47\textwidth}
	\raggedright
		\includegraphics[width=1\linewidth]{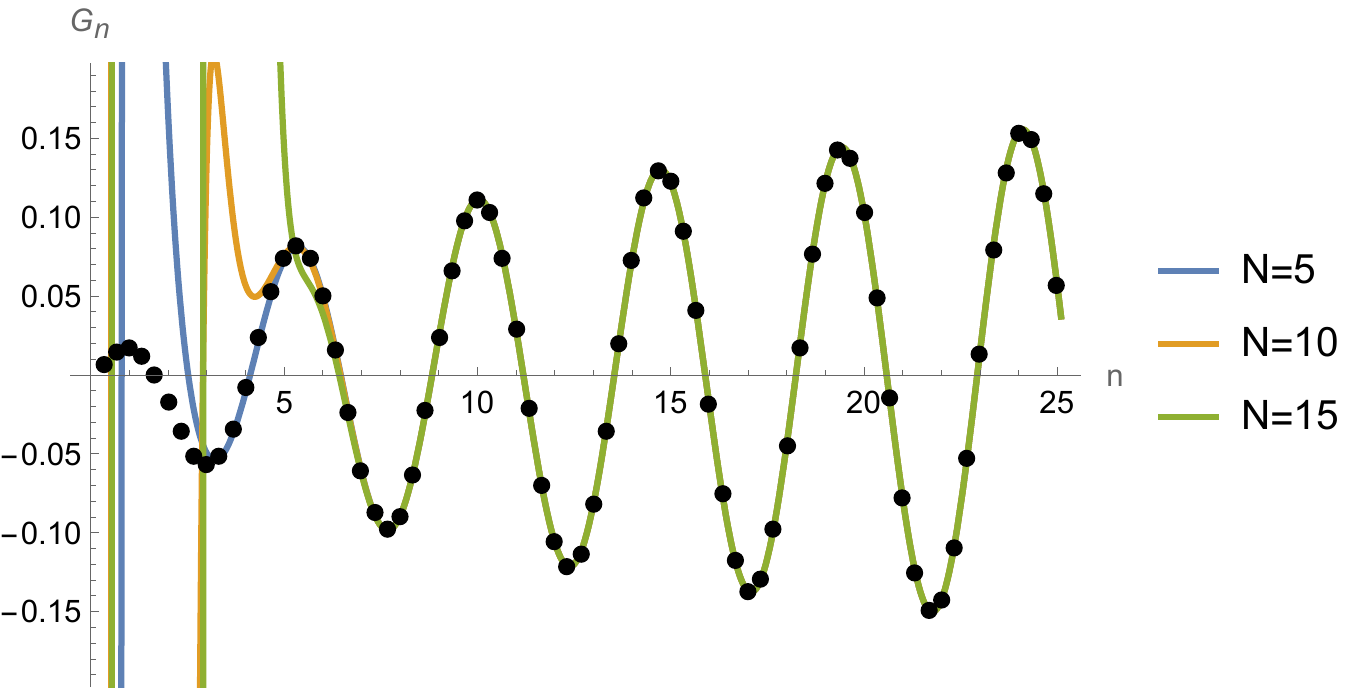}
		\caption{$m=\frac{8}{3},\,n=\frac {p}{3}$}
	\end{subfigure}	
	\begin{subfigure}{0.47\textwidth}
	\raggedright
		\includegraphics[width=1\linewidth]{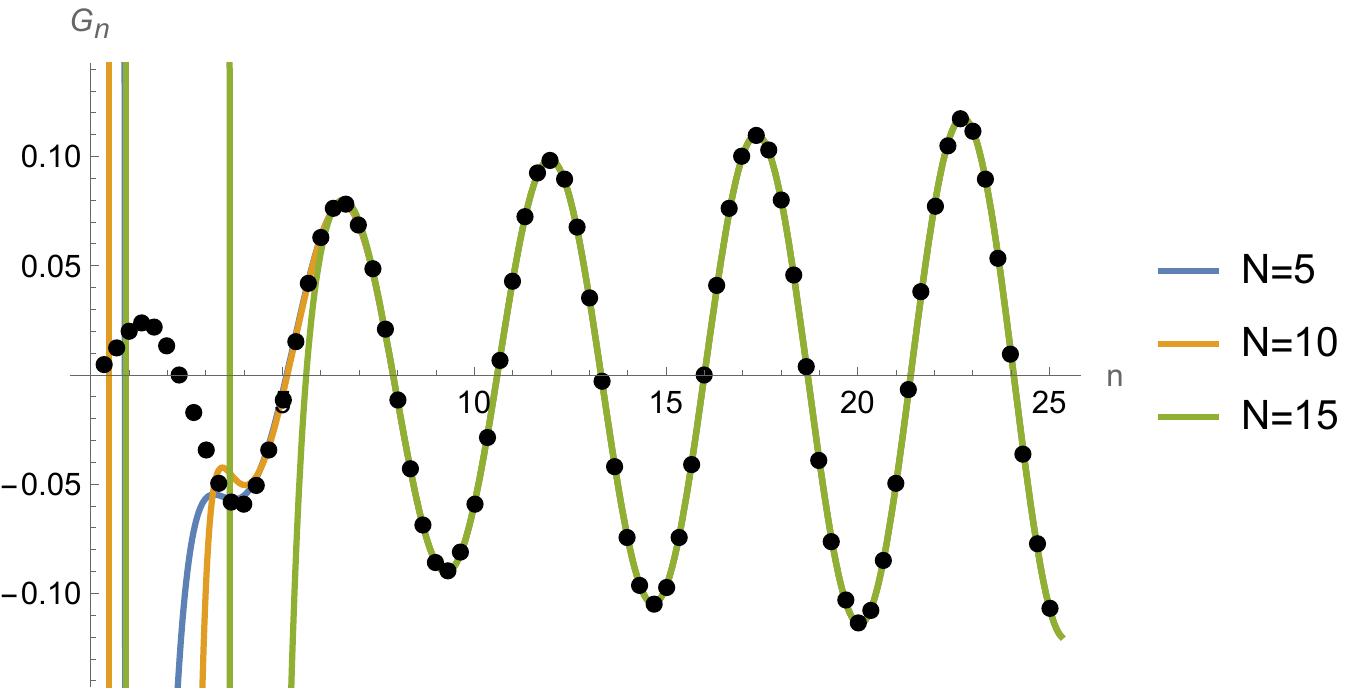}
		\caption{$m=\frac{10}{3},\,n=\frac {p}{3}$}
	\end{subfigure}	
	\qquad
	\begin{subfigure}{0.47\textwidth}
	\raggedright
		\includegraphics[width=1\linewidth]{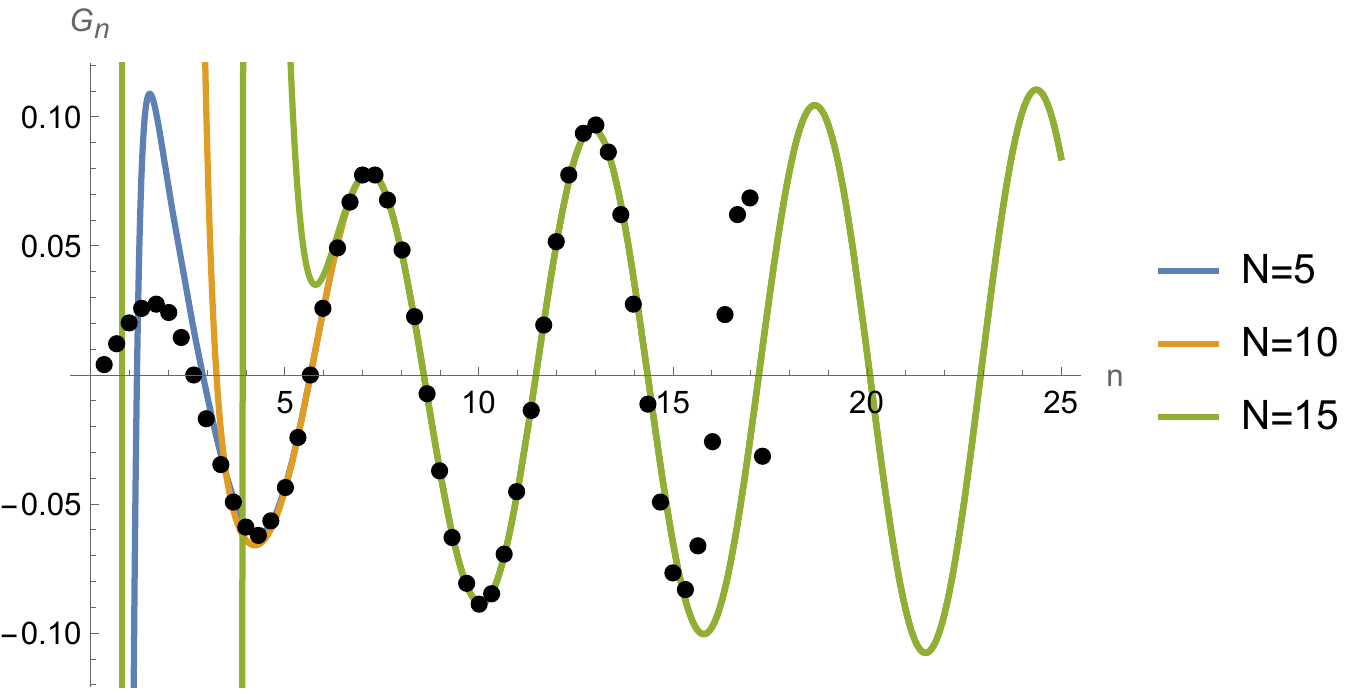}
		\caption{$m=\frac{11}{3},\,n=\frac {p}{3}$}
	\end{subfigure}	
		\caption{Comparison between the asymptotic $1/n$ expansion and the wave function results for $G_n=\<(i\phi)^n\>$. 
		We focus on the $\mathcal {PT}$ symmetric ground states of the non-Hermitian Hamiltonian $H=p^2-(i\phi)^m$. 
		For the minimally singular solutions, the truncation order of the $1/n$ series is given by $N=5,10,15$. 
		We consider several fractional $m$ examples in the range $2<m<4$, 
		which can be studied by the standard method of diagonalizing the truncated Hamiltonian of size $250\times 250$. 
		We use the approximate ground-state wave functions to compute $G_n$ at various $n$, 
		which are labelled by the integer parameter $p$. 
		The large $n$ expansion results match well with the diagonalization results for sufficiently large $n$. 
		On the other hand, if $n$ is too large, the diagonalization results would have noticeable errors. 
		In the $m=11/3$ case, the wave function results exhibit significant errors for $n>14$ due to the slow convergence of the diagonalization method near $m=4$. 
		}
	\label{fig_ix-m-trajectory}
\end{figure}

In Fig. \ref{fig_ix-m-trajectory}, 
we present the Green's functions at various real $n$ 
for $m=\frac 5 2, \frac 7 2$ and $m=\frac 7 3, \frac 8 3, \frac {10} 3,\frac {11} 3$. 
The results from the ground state wave functions \footnote{The approximate wave function is obtained from the diagonalization of a truncated Hamiltonian in the basis of the harmonic oscillator eigenfunctions.  
For non-integer $m$, the diagonalization method is computationally more expensive as all the matrix elements are nonzero, which is associated with a nonlocal nature. 
For integral $m$, only the near diagonal matrix elements are nonzero.  
As the number of nonzero elements grows with $m$, the integral $m$ diagonalization also becomes more expensive at larger $m$.  
}
are again in perfect agreement with the minimally singular solutions with $a_{-1}=(a_1)^\ast\neq 0$. 
To show the asymptotic nature of the large $n$ expansion, 
we present the results associated with the truncated $1/n$ series of order $n^{-5},n^{-10},n^{-15}$. 
We do not use the recursion relation \eqref{ixm-recurion} to improve the low $n$ results 
as in the quartic example in Fig. \ref{fig_x4-trajectory-fixed-Im(n)} 
and the cubic example in Fig. \ref{fig_ix-3-b}. 
As expected, a higher order $1/n$ series gives more accurate estimates at large $n$, 
but the asymptotic series becomes less reliable at small $n$ as the truncation order $N$ increases. 

In some sense, a fractional $m$ case can be viewed 
as a multi-fold covering version of an integral $m$ case.
For integral $m$, 
the recursion relation \eqref{ixm-recurion} has a $(m+2)$-fold rotational symmetry \eqref{PT-rotation-symmetry-integral} for integral $n$, 
so the number of independent $1/n$ series is $m+2$, 
which also determines the number of initial conditions up to some additional constraints.  
For fractional $m$, the recursion relation \eqref{ixm-recurion} is also invariant 
under the discrete rotation \eqref{PT-rotation-symmetry-integral} for integral $n$, 
but the discrete rotational symmetry is not of the $(m+2)$th order, since $m$ is not an integer. 
A multi-fold covering is needed. 
The degree of the covering is associated with the denominator of $m$, 
while the length of a complete period is related to the numerator of $m+2$. 
If $m=p_1/p_2$ and $(p_1,p_2)$ are integers with no common divisor, 
then the total number of free parameters is $p_1+2p_2-1$. 
Note that $p_1+2p_2$ is the numerator of $m+2$. 
The general form of the large $n$ expansion is almost the same as the integral $m$ case \eqref{large-n-ixm}, 
but the spacing of $j$ is reduced to $1/p_2$. 
The matching procedure is the same as before. 
The analytic expression for $\arg[a_1]$ in \eqref{PT-ratio} also applies to the fractional $m$ cases.

For illustration, let us consider the concrete example of $m=5/2$, so we have $p_1=5, p_2=2$. 
In this case, the concrete recursion relation \eqref{ixm-recurion} becomes
\be
(n+1)_3G_n-4E(n+3)G_{n+2}=2(2n+17/2)G_{n+9/2}\,, 
\label{recursion-5/2}
\ee
which clearly suggests an extension to half-integral $n$. 
At large $n$, the perturbative $1/n$ series from \eqref{recursion-5/2}  reads
\be
G_n&\sim&\left(\frac 9 4\right)^{\frac{4n}{9}}\left[\G\left(\frac {2n} {9}\right)\right]^2
n^{\frac{7}{18}}
\sum_{k=0,\pm1,\pm2,\dots} a_k\,e^{2\pi i\frac{2kn}{9}}
\left(1+\sum_{j=\frac 1 2,1,\dots, \frac {9N}{2}}
c_{k,j}\left(\frac n 2\right)^{-\frac{2j}{9}}\right)\,
\qquad (n\rightarrow \infty)\,,
\ee
where the $1/n$ series are truncated to order $n^{-N}$. 
Note that $j$ runs over half-integers, as the spacing for $j$ is given by $ 1/{p_2}=1/2$. 
Some explicit coefficients are 
\be
c_{0,\frac 1 2}=4E\,,\quad c_{0,1}=8E^2\,,\quad c_{0,\frac 3 2}=\frac{32}{3}E^3\,,\quad c_{0,2}=\frac {32}{3}E^4
\,,\quad c_{0,\frac 5 2}=\frac{128}{15}E^5\,,\quad c_{0,3}=\frac {256}{45}E^6.
\ee
Since \eqref{recursion-5/2} is invariant under the rotation
\be
G_n\rightarrow G_n\,e^{2\pi i\frac{2kn}{9}}\,,\quad
E\rightarrow E\,e^{2\pi i\frac{(-4k)}{9}}\,,
\ee
the series coefficients satisfy
\be
c_{k,j}[E]=c_{0,j}\big[Ee^{2\pi i\frac{4k}{9}}\big]\,.
\ee
In accordance with a two-fold covering due to $p_2=2$, the number of independent $1/n$ series is $2(m+2)=9$ for half-integral $n$, which is associated with $e^{4\pi i n}=1$, such as $G_1=G_1e^{2\pi i \frac{2k}{9}}$ with $k=9$. 
After imposing the normalization condition $G_0=1$, we can choose the set of free parameters as
$(E,\, G_{1/2},\,G_1,\,G_2,G_3,G_4)$, together with the nonvanishing prefactors $a_1=(a_{-1})^\ast$ of the $1/n$ series. 
\footnote{Note that the recursion relation \eqref{recursion-5/2} implies $G_{3/2}=0, G_{5/2}=-\frac 4 9 E, G_{7/2}=-\frac 8 {13}E G_1$, so we do not choose them. We could replace $G_{1/2}$ with $G_5$ because $G_5=\frac 7 {1368}(64E^2+135G_{1/2})$. }
The total number of free parameters is compatible with the general formula $p_1+2p_2-1=8$. 
The results in Fig. \ref{subfig:trajectory-5-2} are derived from the ground state solution of the matching conditions 
\be
G_n^{(\text{n.p.})}=G_n^{(\text{p.})}\,,\quad
n=M,M+\frac 1 2,\dots,M+\frac 7 2\,,
\ee
with $(M,N)=(50,5), (100,10), (150,15)$. 

If $m$ is irrational, we could consider a rational approximation for $m$, 
then we can again use the fractional $m$ procedure.  
The results should converge to those of the irrational case 
as we improve the rational approximation for $m$.  
Below, we will not use the rational approximation trick 
and study the irrational situation directly.

\subsection{Irrational powers}
\label{sec: irrational}
In the end, let us consider the recursion relation \eqref{ixm-recurion} with irrational $m$. 
As in the fractional $m$ case, the general form of the $1/n$ series is almost identical to \eqref{large-n-ixm}, except for the spacing of $j$.  
Since the series coefficients are given by polynomials in $E$ with increasing degrees, 
it is natural to organize the $1/n$ series by the power of $E$. 
When $m$ is irrational, the powers of the $1/n$ corrections exhibit a clear correlation with the powers of $E$. 
As a result, the subleading terms are encoded in a double summation of $\left(n/2\right)^{-\frac{j}{m+2}}$ with
\be
j=(m-2)j_1+(m+2)j_2\,,\quad
j_1,j_2=0,1,2,3,\dots,
\label{irrational-j}
\ee
where an additional expansion is related to a small $E$ expansion.  
To see the double expansion structure, 
let us rewrite the recursion relation \eqref{ixm-recurion} as
\be
(n+1)_3G_n-2(2n+m+6)G_{n+m+2}=4E(n+3)G_{n+2}\,. 
\label{ixm-recurion-E} 
\ee
At leading order in small $E$, we neglect the $E$ term on the right hand side of \eqref{ixm-recurion-E}, 
then the general solution reads
\be
G_n\sim\left(\frac m 2+1\right)^{\frac{2n}{m+2}}
\frac{\G(\frac {n+1} {m+2})\G(\frac {n+2} {m+2})\G(\frac {n+3} {m+2})}
{(m+2)^{\frac{m-6}{2(m+2)}}\G(\frac {m+2n+6} {2(m+2)})}
\sum_{k=0,\pm1,\pm2,\dots} \,a_k\,e^{2\pi i\frac{kn}{m+2}}
\quad (E\rightarrow 0)\,,
\ee
which is exact if $E=0$.  
Then we can compute the large $n$ expansion systematically at order $E^0$.  
The corresponding $1/n$ series is an integer power series
\be
\frac{\G(\frac {n+1} {m+2})\G(\frac {n+2} {m+2})\G(\frac {n+3} {m+2})}
{(m+2)^{\frac{m-6}{2(m+2)}}\G(\frac {m+2n+6} {2(m+2)})}
&\sim&
\left[\G\left(\frac n {m+2}\right)\right]^2\,n^{-\frac{m-6}{2(m+2)}}
\left(1+\frac{m^2-20m-4}{8(m+2)}\frac 1 n+\dots
\right)\quad
(n\rightarrow \infty)\,, 
\label{large-n-E0}
\ee 
which also encodes the leading term at large $n$ because the $E$ term in \eqref{ixm-recurion-E}  is also subleading at large $n$.
To determine the first order solution in small $E$, 
we substitute the zeroth order solution into the right hand side of \eqref{ixm-recurion-E}
\be
(n+1)_3G_n-2(2n+m+6)G_{n+m+2}
\sim 4E(n+3)
\left(\frac m 2+1\right)^{\frac{2(n+2)}{m+2}}
\frac{\G(\frac {n+3} {m+2})\G(\frac {n+4} {m+2})\G(\frac {n+5} {m+2})}
{(m+2)^{\frac{m-6}{2(m+2)}}\G(\frac {m+2n+10} {2(m+2)})}
\sum_{k} \,a_k\,e^{2\pi i\frac{kn}{m+2}}\,,\qquad
\ee
which is more difficult to solve exactly, 
but we can use the large $n$ expansion \eqref{large-n-E0} to simplify the problem. 
Then we can also compute the $1/n$ series systematically at order $\mathcal O(E^1)$. 
We notice that the powers of $n$ for the $E^1$ terms are different from those of the $E^0$ terms by $-\frac{m-2}{m+2}$. 
We can repeat this procedure to derive the higher order terms in $E$.  
The general form of the $1/n$ series reads
\be
G_n&\sim&
\left(\frac m 2+1\right)^{\frac{2n}{m+2}}\left[\G\left(\frac n {m+2}\right)\right]^2
n^{-\frac{m-6}{2(m+2)}}
\sum_{k=0}^{m+1} a_k\,e^{2\pi i\frac{kn}{m+2}}
\sum_{j_1,j_2}\tilde c_{j_1,j_2}\left(Ee^{2\pi i\frac{2k}{m+2}}\left(\frac n 2\right)^{-\frac{m-2}{m+2}}\right)^{j_1}\left(\frac n 2\right)^{-j_2}\,\quad (n\rightarrow \infty)\,,\nn
\label{small-E-large-n}
\ee
where the double expansion with $j_1,j_2=0,1,2,3,\dots$ is mentioned earlier in \eqref{irrational-j}, 
and the truncation to order $n^{-N}$ is set by $\frac{m-2}{m+2}j_1+j_2\leq N$. 
Note that $j_1$ is related to the small $E$ expansion, which is automatically truncated by the large $n$ expansion. 
Some low order coefficients are
\be
&&\tilde c_{0,0}=1\,,\quad
\tilde c_{0,1}=\frac{m^2-20m-4}{16(m+2)}\,,
\ee
\be
\tilde c_{1,0}=\frac{2}{m-2}\,,\quad
\tilde c_{1,1}=\frac{5m^2-44m+28}{8(m-2)(m+2)}\,,
\ee
\be
\tilde c_{2,0}=\frac{2}{(m-2)^2}\,,\quad
\tilde c_{2,1}=\frac{35 m^3 - 270 m^2 + 412 m - 184}{8(m-2)^2(m+2)(3m-2)}\,,
\ee
which agree with the general forms in \eqref{coefficient-1} and \eqref{coefficient-2} for large $m$. 
In fact, the large $n$ expansion can be computed more systematically for irrational $m$, 
as some rational $m$ coefficients may correspond to degenerate limits and require more case. 
It is more clear why the leading correction is of order $n^{-\frac{m-2}{m+2}}$. 

Below we again consider the minimally singular solution with two nonzero prefactors $a_{-1}=(a_1)^\ast$. 
When $m$ is irrational, the integer $n$ Green's functions are related to the irrational cases by 
the recursion relation \eqref{ixm-recurion-E}.   
A minimal set of $G_n$ is labeled by two integers $(q_1, q_2)$ as
\be
n_{q_1,q_2}=q_1m+q_2\,,
\ee
and the number of independent $G_n$ is infinite. 
We can use the recursion relation \eqref{ixm-recurion-E} to express 
$G_{q_1 m+q_2}$ in terms of $G_{q_1 m+q_2-2}$ and $G_{(q_1+1)m+q_2}$, 
so we have at most two free parameters for each $q_1$, but there are infinitely many $q_1$.

If $m$ is rational, the number of independent $q_1$ is reduced to a finite number, 
which is related to the denominator of $m$ or the degree of covering, 
so only finitely many $G_n$ in the minimal set are independent parameters for rational $m$. 
An irrational $m$ case can be viewed as a multiple covering of infinite degree.  
In other words, the number of free parameters is infinite due to absence of periodicity conditions. 
Although there are infinitely many independent parameters for irrational $m$, 
we can still truncate the system of self-consistent equations and derive accurate solutions 
from the corresponding matching conditions.  
The number of independent parameters and the accuracy of the determinations increase with the size of the truncated system. 

As an explicit example, we consider the irrational case 
\be
m=\pi=3.14159265\dots\,,
\ee
so the explicit recursion relation reads
\be
(n+1)_3G_n-4E(n+3)G_{n+2}=2(2n+\pi+6)G_{n+\pi+2}\,.
\label{ixm-recurion-pi}
\ee
For concreteness, the truncation parameters are set to be $(M,N)=(50,5)$.   
We need to express the Green's functions in the minimal set with $0\leq n\leq 50$ in terms of 
those with $n>50$ and an independent set of free parameters. 
We choose the recursion relation \eqref{ixm-recurion-pi} with $n=q_1 \pi+q_2$, 
where $(q_1, q_2)$ are non-negative integers satisfying $0\leq q_1< 50/\pi$ and $0\leq q_2\leq 48-q_1\pi$. 
The independent parameters for 
the Green's functions with $n\leq 50$ are associated with
$n=0,1,\pi,\pi+1,\dots,14\pi,14\pi+1, 15\pi$,
as $15\pi+1\approx 48.1>48$ is greater than $M-2$ and $16\pi \approx 50.3>50$ is greater than $M$. 
The fact that the number of free parameters in the truncated system grows with the matching parameter $M$ 
is consistent with the existence of infinitely many free parameters for irrational $m$. 
The Green's functions  $G_{n>50}$ with $n=q_1\pi+q_2$ are associated with $1\leq q_1\leq 16$ and $50<n \leq 50+\pi$, such as $n=\pi+47,\pi+48,\pi+49,\pi+50, 2\pi+44,2\pi+45,2\pi+46,\dots$. 
The recursion relation \eqref{ixm-recurion-pi} at the special values $n=-2,-1$ 
eliminates two more free parameters. 
The normalization condition also fixes $G_0=1$. 
There are 2 more free parameters in the $1/n$ series \eqref{small-E-large-n}, i.e., $\text{Re}[a_1]$ and $\text{Im}[a_1]$, 
together with the energy $E$.  
Therefore, the total number of free parameters is $31-3+3=31$ and we need $31$ matching conditions. 
We choose the largest 31 cases in the minimal set with $n\leq 50$, 
i.e., $n=50,7\pi+28,14\pi+6,6\pi+31,\dots, \pi+45, 8\pi+23$. 
The Green's functions with $n>50$ are directly approximated by the $1/n$ series \eqref{small-E-large-n}. 
\footnote{We could also interpret them as free parameters, but their matching conditions do not involve the small $n$ parameters. Their existence is related to the absence of periodicity conditions for irrational $m$, 
which is a sharp difference from the rational-$m$ situation. }  
Then the minimization of the $\eta$ function \eqref{eta-HO} gives an estimate for the ground state energy
$E_0\approx 1.1940021$,
which is well consistent with the diagonalization result $E_0=1.19400128\dots$.
Although the number of free parameters is significantly greater than the rational examples 
\footnote{In this irrational example, we set a higher working precision for numerical stability. }, 
the error in the ground state energy $\D E_0\approx 8\times 10^{-7}$ is still compatible with the precision pattern in Fig. \ref{fig_ix-m-error-b}. 
\footnote{The error here is expected to be slightly greater, as the matching condition is evaluated at $n\leq M$, while the integral and fractional cases are evaluated at $n\geq M$. }
The argument of $a_1$ in this irrational case is also consistent with 
the analytic expression \eqref{PT-ratio} with a numerical error of the same order as $\D E_0$.

\section{Discussion}
\label{sec:discussion}
In this work, we have further investigated the bootstrap approach 
based on the analytic continuation of 
the Green's functions $\<\phi^n\>$ or $\<(i\phi)^n\>$ to complex $n$. 
We used the quantum harmonic oscillator to illustrate various aspects of the bootstrap analysis: 
\begin{itemize}
\item The large $n$ expansion of the Green's functions $\<\phi^n\>$.
\item The matching conditions for the non-perturbative finite $n$ solutions and the perturbative large $n$ expansion series. 
\item The principle of minimal singularity as an exact quantization condition. 
\item The high energy asymptotic behavior of the bootstrap solutions. 
\end{itemize}
The two questions raised in the introduction have been addressed: 
\begin{enumerate}
\item
The Green's functions $\<\phi^n\>$ or $\<(i\phi)^n\>$ under consideration are well consistent with the non-integer $n$ results from the standard wave function formulation. 
Furthermore, naively different branches of Green's functions are unified by the $n$ analytic continuation,
such as the cubic example in Fig. \ref{fig_ix-3}. 
\item
The $\phi^n$ trajectory bootstrap method gives highly accurate results 
for the anharmonic potentials $V(\phi)=\phi^2+\phi^{m}$ and $V(\phi)=-(i\phi)^{m}$ 
in a large range of integral $m$. 
For the $\mathcal {PT}$ invariant potential, we also successfully obtain accurate bootstrap solutions for fractional and irrational $m$. 
\end{enumerate}

For $D=0$, the basic physical observables in scalar field theories are associated with the composite operators $\phi^n$. 
\footnote{In \cite{Li:2024ggr}, we revisit the matrix models by the analytic continuation of $\<\text{Tr} M^n\>$ in $n$, 
where $M$ represent a random matrix.  
For the two-matrix model with $\text{Tr}[A, B]^2$ interaction and quartic potentials, 
we obtain highly accurate results at a relatively low computational cost. 
See also \cite{Li:2023tic} for a recent conformal bootstrap study of the 3D Ising model based on analytic trajectories.  }
It is natural to consider the analytic continuation in the number of fundamental fields. 
For $D\geq 1$, we can construct composite operators using 
both fundamental fields and derivatives, 
so it is natural to complexify the numbers of derivatives as well. 
\footnote{The number of free derivative indices is related to angular momentum, 
so this can be viewed as a generalization of the Regge trajectory. 
We may also consider the analytic continuation of various symmetry representations. }
In the Hamiltonian formulation, we can consider both the fundamental fields and their canonical conjugates. 
It is also interesting to go beyond the one-point functions of composite operators. 
For applications to particle physics and condensed matter physics,  
it is important to introduce the fermionic degrees of freedom.  
We may analytically continue the numbers of bilinear operators, such as $(\bar\psi\psi)^n$. 
For a periodic system, we can also analytically continue the exponents of exponential operators 
and the quasi-momentum to bootstrap the band structure. 
\footnote{For instance, the self-consistency equation for $\<e^{inx}\>$ in \cite{Tchoumakov:2021mnh} 
can be solved systematically by the large $n$ expansion
\be
\<e^{inx}\>=\<\cos(nx)\>
\sim a_0(-4)^{n}\frac{(3/2)_{n-1}}{\G(n+1)^3}
\left(1-\frac {4E}{n}
+\dots\right)
+a_1(-4)^{-n}\frac{\G(n)^3}{(3/2)_{n-1}}
\left(1+\frac {4E}{n}
+\dots\right)\,,\quad (n\rightarrow \infty)
\ee 
where the Hamiltonian is $H=p^2+2\cos(x)$. 
The results for $E$ and $\<\cos(x)\>$ are consistent with those from the positivity constraints. 
It is also interesting to consider the momentum counterpart, 
such as the analytic continuation of $\<e^{inp}\>$ in $n$. 
To further deduce the Bloch band spectrum, it is useful to consider the analytic continuation in the Bloch momentum. 
}

It would be fascinating to further explore the various possibilities 
on the interplay between self-consistency and analyticity.
\footnote{A natural starting point is to consider the $n$ analytic continuation in (generalized) free CFTs. 
See \cite{Caron-Huot:2017vep,Simmons-Duffin:2017nub,Kravchuk:2018htv} for the application of analyticity in spin to the conformal bootstrap. 
The analytic continuation in spin is closely related to the nonlocal light-ray operators. 
Similarly, the analytic continuation in $n$  involves the nonlocal properties of the local observables. 
Note that a non-integer power of the position operator in the momentum representation 
is associated with fractional calculus. 
} 
We hope that this will lead to 
efficient non-perturbative methods for 
studying the strongly coupled or strongly correlated physics 
in more realistic quantum field theories and quantum many-body systems. 

\section*{Acknowledgments}
I would like to thank the referees for the insightful comments and constructive suggestions. 
This work was supported by 
the Natural Science Foundation of China (Grant No. 12205386) and 
the Guangzhou Municipal Science and Technology Project (Grant No. 2023A04J0006).


\begin{thebibliography}{10}
\bibitem{Rattazzi:2008pe}
R.~Rattazzi, V.~S.~Rychkov, E.~Tonni and A.~Vichi,
``Bounding scalar operator dimensions in 4D CFT,''
JHEP \textbf{12}, 031 (2008)
doi:10.1088/1126-6708/2008/12/031
[arXiv:0807.0004 [hep-th]].

\bibitem{Poland:2018epd}
D.~Poland, S.~Rychkov and A.~Vichi,
``The Conformal Bootstrap: Theory, Numerical Techniques, and Applications,''
Rev. Mod. Phys. \textbf{91}, 015002 (2019)
doi:10.1103/RevModPhys.91.015002
[arXiv:1805.04405 [hep-th]].

\bibitem{Dyson:1949ha}
F.~J.~Dyson,
``The S matrix in quantum electrodynamics,''
Phys. Rev. \textbf{75}, 1736-1755 (1949)
doi:10.1103/PhysRev.75.1736

\bibitem{Schwinger:1951ex}
J.~S.~Schwinger,
``On the Green's functions of quantized fields. 1.,''
Proc. Nat. Acad. Sci. \textbf{37}, 452-455 (1951)
doi:10.1073/pnas.37.7.452

\bibitem{Schwinger:1951hq}
J.~S.~Schwinger,
``On the Green's functions of quantized fields. 2.,''
Proc. Nat. Acad. Sci. \textbf{37}, 455-459 (1951)
doi:10.1073/pnas.37.7.455

\bibitem{Makeenko:1979pb}
Y.~M.~Makeenko and A.~A.~Migdal,
``Exact Equation for the Loop Average in Multicolor QCD,''
Phys. Lett. B \textbf{88}, 135 (1979)
[erratum: Phys. Lett. B \textbf{89}, 437 (1980)]
doi:10.1016/0370-2693(79)90131-X

\bibitem{Makeenko:1980vm}
Y.~Makeenko and A.~A.~Migdal,
``Quantum Chromodynamics as Dynamics of Loops,''
Sov. J. Nucl. Phys. \textbf{32}, 431 (1980)
doi:10.1016/0550-3213(81)90258-3

\bibitem{Migdal:1983qrz}
A.~A.~Migdal,
``Loop Equations and 1/N Expansion,''
Phys. Rept. \textbf{102}, 199-290 (1983)
doi:10.1016/0370-1573(83)90076-5

\bibitem{Anderson:2016rcw}
P.~D.~Anderson and M.~Kruczenski,
``Loop Equations and bootstrap methods in the lattice,''
Nucl. Phys. B \textbf{921} (2017), 702-726
[arXiv:1612.08140 [hep-th]].

\bibitem{Lin:2020mme}
H.~W.~Lin,
``Bootstraps to strings: solving random matrix models with positivity,''
JHEP \textbf{06} (2020), 090
[arXiv:2002.08387 [hep-th]].

\bibitem{Hessam:2021byc}
H.~Hessam, M.~Khalkhali and N.~Pagliaroli,
``Bootstrapping Dirac ensembles,''
J. Phys. A \textbf{55}, no.33, 335204 (2022)
[arXiv:2107.10333 [hep-th]].

\bibitem{Kazakov:2021lel}
V.~Kazakov and Z.~Zheng,
``Analytic and numerical bootstrap for one-matrix model and \textquotedblleft{}unsolvable\textquotedblright{} two-matrix model,''
JHEP \textbf{06} (2022), 030
[arXiv:2108.04830 [hep-th]].

\bibitem{Kazakov:2022xuh}
V.~Kazakov and Z.~Zheng,
``Bootstrap for lattice Yang-Mills theory,''
Phys. Rev. D \textbf{107} (2023) no.5, L051501
[arXiv:2203.11360 [hep-th]].

\bibitem{Cho:2022lcj}
M.~Cho, B.~Gabai, Y.~H.~Lin, V.~A.~Rodriguez, J.~Sandor and X.~Yin,
``Bootstrapping the Ising Model on the Lattice,''
[arXiv:2206.12538 [hep-th]].

\bibitem{Li:2023nip}
W.~Li,
``Taming Dyson-Schwinger Equations with Null States,''
Phys. Rev. Lett. \textbf{131}, no.3, 031603 (2023)
doi:10.1103/PhysRevLett.131.031603
[arXiv:2303.10978 [hep-th]].

\bibitem{Han:2020bkb}
X.~Han, S.~A.~Hartnoll and J.~Kruthoff,
``Bootstrapping Matrix Quantum Mechanics,''
Phys. Rev. Lett. \textbf{125} (2020) no.4, 041601
[arXiv:2004.10212 [hep-th]].

\bibitem{Han}
X. Han,
``Quantum Many-body Bootstrap,''
[arXiv:2006.06002[cond-mat]].

\bibitem{Berenstein:2021dyf}
D.~Berenstein and G.~Hulsey,
``Bootstrapping Simple QM Systems,''
[arXiv:2108.08757 [hep-th]].


\bibitem{Bhattacharya:2021btd}
J.~Bhattacharya, D.~Das, S.~K.~Das, A.~K.~Jha and M.~Kundu,
``Numerical bootstrap in quantum mechanics,''
Phys. Lett. B \textbf{823} (2021), 136785
[arXiv:2108.11416 [hep-th]].

\bibitem{Aikawa:2021eai}
Y.~Aikawa, T.~Morita and K.~Yoshimura,
``Application of bootstrap to a \ensuremath{\theta} term,''
Phys. Rev. D \textbf{105} (2022) no.8, 085017
[arXiv:2109.02701 [hep-th]].

\bibitem{Berenstein:2021loy}
D.~Berenstein and G.~Hulsey,
``Bootstrapping more QM systems,''
J. Phys. A \textbf{55} (2022) no.27, 275304
[arXiv:2109.06251 [hep-th]].

\bibitem{Tchoumakov:2021mnh}
S.~Tchoumakov and S.~Florens,
``Bootstrapping Bloch bands,''
J. Phys. A \textbf{55} (2022) no.1, 015203
[arXiv:2109.06600 [cond-mat.mes-hall]].


\bibitem{Aikawa:2021qbl}
Y.~Aikawa, T.~Morita and K.~Yoshimura,
``Bootstrap method in harmonic oscillator,''
Phys. Lett. B \textbf{833} (2022), 137305
[arXiv:2109.08033 [hep-th]].

\bibitem{Du:2021hfw}
B.~n.~Du, M.~x.~Huang and P.~x.~Zeng,
``Bootstrapping Calabi\textendash{}Yau quantum mechanics,''
Commun. Theor. Phys. \textbf{74} (2022) no.9, 095801
[arXiv:2111.08442 [hep-th]].

\bibitem{Lawrence:2021msm}
S.~Lawrence,
``Bootstrapping Lattice Vacua,''
[arXiv:2111.13007 [hep-lat]].

\bibitem{Bai:2022yfv}
D.~Bai,
``Bootstrapping the deuteron,''
[arXiv:2201.00551 [nucl-th]].

\bibitem{Nakayama:2022ahr}
Y.~Nakayama,
``Bootstrapping microcanonical ensemble in classical system,''
Mod. Phys. Lett. A \textbf{37} (2022) no.09, 2250054
[arXiv:2201.04316 [hep-th]].

\bibitem{Li:2022prn}
W.~Li,
``Null bootstrap for non-Hermitian Hamiltonians,''
Phys. Rev. D \textbf{106}, no.12, 125021 (2022)
doi:10.1103/PhysRevD.106.125021
[arXiv:2202.04334 [hep-th]].

\bibitem{Khan:2022uyz}
S.~Khan, Y.~Agarwal, D.~Tripathy and S.~Jain,
``Bootstrapping PT symmetric quantum mechanics,''
Phys. Lett. B \textbf{834} (2022), 137445
[arXiv:2202.05351 [quant-ph]].

\bibitem{Berenstein:2022ygg}
D.~Berenstein and G.~Hulsey,
``Anomalous bootstrap on the half-line,''
Phys. Rev. D \textbf{106}, no.4, 045029 (2022)
[arXiv:2206.01765 [hep-th]].

\bibitem{Morita:2022zuy}
T.~Morita,
``Universal bounds on quantum mechanics through energy conservation and the bootstrap method,''
PTEP \textbf{2023} (2023) no.2, 023A01
[arXiv:2208.09370 [hep-th]].

\bibitem{Blacker:2022szo}
M.~J.~Blacker, A.~Bhattacharyya and A.~Banerjee,
``Bootstrapping the Kronig-Penney model,''
Phys. Rev. D \textbf{106} (2022) no.11, 11
[arXiv:2209.09919 [quant-ph]].

\bibitem{Berenstein:2022unr}
D.~Berenstein and G.~Hulsey,
``Semidefinite programming algorithm for the quantum mechanical bootstrap,''
Phys. Rev. E \textbf{107} (2023) no.5, L053301
[arXiv:2209.14332 [hep-th]].

\bibitem{Nancarrow:2022wdr}
C.~O.~Nancarrow and Y.~Xin,
``Bootstrapping the gap in quantum spin systems,''
JHEP \textbf{08}, 052 (2023)
doi:10.1007/JHEP08(2023)052
[arXiv:2211.03819 [hep-th]].

\bibitem{Lawrence:2022vsb}
S.~Lawrence,
``Semidefinite programs at finite fermion density,''
Phys. Rev. D \textbf{107} (2023) no.9, 094511
[arXiv:2211.08874 [hep-lat]].

\bibitem{Lin:2023owt}
H.~W.~Lin,
``Bootstrap bounds on D0-brane quantum mechanics,''
JHEP \textbf{06} (2023), 038
[arXiv:2302.04416 [hep-th]]. 

\bibitem{Guo:2023gfi}
Y.~Guo and W.~Li,
``Solving anharmonic oscillator with null states: Hamiltonian bootstrap and Dyson-Schwinger equations,''
Phys. Rev. D \textbf{108}, no.12, 125002 (2023)
doi:10.1103/PhysRevD.108.125002
[arXiv:2305.15992 [hep-th]].

\bibitem{Berenstein:2023ppj}
D.~Berenstein and G.~Hulsey,
``One-dimensional reflection in the quantum mechanical bootstrap,''
Phys. Rev. D \textbf{109}, no.2, 025013 (2024)
doi:10.1103/PhysRevD.109.025013
[arXiv:2307.11724 [hep-th]].

\bibitem{Fan:2023bld}
W.~Fan and H.~Zhang,
``Non-perturbative instanton effects in the quartic and the sextic double-well potential by the numerical bootstrap approach,''
[arXiv:2308.11516 [hep-th]].


\bibitem{Li:2023ewe}
W.~Li,
``Principle of minimal singularity for Green\textquoteright{}s functions,''
Phys. Rev. D \textbf{109}, no.4, 045012 (2024)
doi:10.1103/PhysRevD.109.045012
[arXiv:2309.02201 [hep-th]].


\bibitem{John:2023him}
R.~R.~John and K.~P.~R,
``Anharmonic oscillators and the null bootstrap,''
[arXiv:2309.06381 [quant-ph]].

\bibitem{Fan:2023tlh}
W.~Fan, H.~Zhang and Z.~Li,
``Unify the Effect of Anharmonicity in Double-Wells and Anharmonic Oscillators,''
Int. J. Theor. Phys. \textbf{63}, no.10, 266 (2024)
doi:10.1007/s10773-024-05774-w
[arXiv:2309.09269 [quant-ph]].

\bibitem{Guo:2023qtt}
Y.~Guo and W.~Li,
``Anomalous dimensions of partially conserved higher-spin currents from conformal field theory: Bosonic \ensuremath{\phi}2n theories,''
Phys. Rev. D \textbf{109}, no.2, 025015 (2024)
doi:10.1103/PhysRevD.109.025015
[arXiv:2305.16916 [hep-th]].

\bibitem{Guo:2024bll}
Y.~Guo and W.~Li,
``Anomalous dimensions from conformal field theory: generalized $\phi^{2n+1}$ theories,''
[arXiv:2408.12344 [hep-th]].

\bibitem{Bender:1988bp}
C.~M.~Bender, F.~Cooper and L.~M.~Simmons,
``Nonunique Solution to the Schwinger-dyson Equations,''
Phys. Rev. D \textbf{39}, 2343-2349 (1989)
doi:10.1103/PhysRevD.39.2343

\bibitem{Bender:2022eze}
C.~M.~Bender, C.~Karapoulitidis and S.~P.~Klevansky,
``Underdetermined Dyson-Schwinger Equations,''
Phys. Rev. Lett. \textbf{130}, no.10, 101602 (2023)
doi:10.1103/PhysRevLett.130.101602
[arXiv:2211.13026 [math-ph]].


\bibitem{Bender:2023ttu}
C.~M.~Bender, C.~Karapoulitidis and S.~P.~Klevansky,
``Dyson-Schwinger equations in zero dimensions and polynomial approximations,''
Phys. Rev. D \textbf{108}, no.5, 056002 (2023)
doi:10.1103/PhysRevD.108.056002
[arXiv:2307.01008 [math-ph]].

\bibitem{Regge:1959mz}
T.~Regge,
``Introduction to complex orbital momenta,''
Nuovo Cim. \textbf{14}, 951 (1959)
doi:10.1007/BF02728177

\bibitem{Chew:book}
G.~F.~Chew, 
``The S-matrix theory of strong interactions,'' 
(W. A. Benjamin, Inc., New York, 1961).

\bibitem{Chew:1961ev}
G.~F.~Chew and S.~C.~Frautschi,
``Principle of Equivalence for All Strongly Interacting Particles Within the S Matrix Framework,''
Phys. Rev. Lett. \textbf{7}, 394-397 (1961)
doi:10.1103/PhysRevLett.7.394

\bibitem{Bender:1998ke}
C.~M.~Bender and S.~Boettcher,
``Real spectra in nonHermitian Hamiltonians having PT symmetry,''
Phys. Rev. Lett. \textbf{80}, 5243-5246 (1998)
doi:10.1103/PhysRevLett.80.5243
[arXiv:physics/9712001 [physics]].

\bibitem{Bender:1999ek}
C.~M.~Bender, K.~A.~Milton and V.~Savage,
``Solution of Schwinger-Dyson equations for PT symmetric quantum field theory,''
Phys. Rev. D \textbf{62}, 085001 (2000)
doi:10.1103/PhysRevD.62.085001
[arXiv:hep-th/9907045 [hep-th]].

\bibitem{Bender:2007nj}
C.~M.~Bender,
``Making sense of non-Hermitian Hamiltonians,''
Rept. Prog. Phys. \textbf{70}, 947 (2007)
doi:10.1088/0034-4885/70/6/R03
[arXiv:hep-th/0703096 [hep-th]].

\bibitem{Bender:2010hf}
C.~M.~Bender and S.~P.~Klevansky,
``Families of particles with different masses in PT-symmetric quantum field theory,''
Phys. Rev. Lett. \textbf{105}, 031601 (2010)
doi:10.1103/PhysRevLett.105.031601
[arXiv:1002.3253 [hep-th]].

\bibitem{r5} 
C.~M.~Bender {\it et al.}, {\it PT Symmetry: in Quantum and
Classical Physics} (World Scientific, Singapore, 2019).

\bibitem{Bender:2023cem}
C.~M.~Bender and D.~W.~Hook,
``PT-symmetric quantum mechanics,''
[arXiv:2312.17386 [quant-ph]].

\bibitem{vonGehlen:1994rp}
G.~von Gehlen,
``NonHermitian tricriticality in the Blume-Capel model with imaginary field,''
[arXiv:hep-th/9402143 [hep-th]].

\bibitem{Lencses:2022ira}
M.~Lencs\'es, A.~Miscioscia, G.~Mussardo and G.~Tak\'acs,
``Multicriticality in Yang-Lee edge singularity,''
JHEP \textbf{02}, 046 (2023)
doi:10.1007/JHEP02(2023)046
[arXiv:2211.01123 [hep-th]].

\bibitem{Lencses:2023evr}
M.~Lencs\'es, A.~Miscioscia, G.~Mussardo and G.~Tak\'acs,
``$ \mathcal{PT} $ breaking and RG flows between multicritical Yang-Lee fixed points,''
JHEP \textbf{09}, 052 (2023)
doi:10.1007/JHEP09(2023)052
[arXiv:2304.08522 [cond-mat.stat-mech]].

\bibitem{Gang:2023rei}
D.~Gang, H.~Kim and S.~Stubbs,
``Three-Dimensional Topological Field Theories and Nonunitary Minimal Models,''
Phys. Rev. Lett. \textbf{132}, no.13, 131601 (2024)
doi:10.1103/PhysRevLett.132.131601
[arXiv:2310.09080 [hep-th]].


\bibitem{Zamolodchikov:1986db}
A.~B.~Zamolodchikov,
``Conformal Symmetry and Multicritical Points in Two-Dimensional Quantum Field Theory. (In Russian),''
Sov. J. Nucl. Phys. \textbf{44}, 529-533 (1986)

\bibitem{Lencses:2024wib}
M.~Lencs\'es, A.~Miscioscia, G.~Mussardo and G.~Tak\'acs,
``Ginzburg-Landau description for multicritical Yang-Lee models,''
JHEP \textbf{08}, 224 (2024)
doi:10.1007/JHEP08(2024)224
[arXiv:2404.06100 [cond-mat.stat-mech]].


\bibitem{Gliozzi:2013ysa}
F.~Gliozzi,
``More constraining conformal bootstrap,''
Phys. Rev. Lett. \textbf{111}, 161602 (2013)
doi:10.1103/PhysRevLett.111.161602
[arXiv:1307.3111 [hep-th]].

\bibitem{Gliozzi:2014jsa}
F.~Gliozzi and A.~Rago,
``Critical exponents of the 3d Ising and related models from Conformal Bootstrap,''
JHEP \textbf{10}, 042 (2014)
doi:10.1007/JHEP10(2014)042
[arXiv:1403.6003 [hep-th]].

\bibitem{Li:2017agi}
W.~Li,
``Inverse Bootstrapping Conformal Field Theories,''
JHEP \textbf{01}, 077 (2018)
doi:10.1007/JHEP01(2018)077
[arXiv:1706.04054 [hep-th]].

\bibitem{Hikami:2017hwv}
S.~Hikami,
``Conformal bootstrap analysis for the Yang\textendash{}Lee edge singularity,''
PTEP \textbf{2018}, no.5, 053I01 (2018)
doi:10.1093/ptep/pty054
[arXiv:1707.04813 [hep-th]].

\bibitem{Li:2017ukc}
W.~Li,
``New method for the conformal bootstrap with OPE truncations,''
[arXiv:1711.09075 [hep-th]].

\bibitem{Li:2021uki}
W.~Li,
``Ising model close to d=2,''
Phys. Rev. D \textbf{105}, no.9, L091902 (2022)
doi:10.1103/PhysRevD.105.L091902
[arXiv:2107.13679 [hep-th]].

\bibitem{Li:2023tic}
W.~Li,
``Easy bootstrap for the 3D Ising model: a hybrid approach of the lightcone bootstrap and error minimization methods,''
JHEP \textbf{07}, 047 (2024)
doi:10.1007/JHEP07(2024)047
[arXiv:2312.07866 [hep-th]].

\bibitem{Sibuya}
Y. ~Sibuya, ``Global theory of a second-order linear ordinary differential equation with polynomial coefficient", (Amsterdam: North-Holland 1975).

\bibitem{Li:2024ggr}
W.~Li,
``Analytic trajectory bootstrap for matrix models,''
[arXiv:2407.08593 [hep-th]].

\bibitem{Caron-Huot:2017vep}
S.~Caron-Huot,
``Analyticity in Spin in Conformal Theories,''
JHEP \textbf{09}, 078 (2017)
doi:10.1007/JHEP09(2017)078
[arXiv:1703.00278 [hep-th]].

\bibitem{Simmons-Duffin:2017nub}
D.~Simmons-Duffin, D.~Stanford and E.~Witten,
``A spacetime derivation of the Lorentzian OPE inversion formula,''
JHEP \textbf{07}, 085 (2018)
doi:10.1007/JHEP07(2018)085
[arXiv:1711.03816 [hep-th]].

\bibitem{Kravchuk:2018htv}
P.~Kravchuk and D.~Simmons-Duffin,
``Light-ray operators in conformal field theory,''
JHEP \textbf{11}, 102 (2018)
doi:10.1007/JHEP11(2018)102
[arXiv:1805.00098 [hep-th]].


\end{thebibliography}
\end{document}